\documentclass[fleqn,usenatbib]{mnras}

\usepackage{newtxtext,newtxmath}

\usepackage[T1]{fontenc}

\DeclareRobustCommand{\VAN}[3]{#2}
\let\VANthebibliography\thebibliography
\def\thebibliography{\DeclareRobustCommand{\VAN}[3]{##3}\VANthebibliography}

\usepackage{xcolor}
\usepackage{multirow}
\usepackage{soul}
\usepackage{subcaption}

\usepackage{graphicx}	
\usepackage{amsmath}	

\DeclareCaptionFormat{cont}{#1 (cont.)#2#3\par}
\newenvironment{compactlists}
{ \begin{itemize}
    \setlength{\itemsep}{0pt}
    \setlength{\parskip}{0pt}
    \setlength{\parsep}{0pt}     }
{ \end{itemize}                  } 

\title[Diverse quenching pathways of PSB galaxies]{The diverse quenching pathways of post-starburst galaxies in SDSS-IV MaNGA}

\author[H-H. Leung et al.]{
Ho-Hin Leung,$^{1,2}$\thanks{E-mail: hleung2@roe.ac.uk}
Vivienne Wild,$^{1}$
Michail Papathomas,$^{3}$
Adam C. Carnall,$^{2}$
and Yanmei Chen$^{4,5}$
\\
$^{1}$SUPA\footnote{Scottish Universities Physics Alliance}, School of Physics \& Astronomy, 
University of St Andrews, 
North Haugh, St Andrews, Fife KY16 9SS, UK\\
$^{2}$SUPA, Institute for Astronomy, University of Edinburgh, Royal Observatory, Edinburgh EH9 3HJ, UK\\
$^{3}$School of Mathematics and Statistics, University of St Andrews, 
North Haugh, St Andrews, 
Fife, KY16 9SS, UK\\
$^{4}$School of Astronomy and Space Science, Nanjing University, Nanjing 210093, China \\
$^{5}$Key Laboratory of Modern Astronomy and Astrophysics (Nanjing University), Ministry of Education, Nanjing 210023, China
}

\date{Accepted XXX. Received YYY; in original form ZZZ}

\pubyear{\the\year{}}

\begin{document}
\label{firstpage}
\pagerange{\pageref{firstpage}--\pageref{lastpage}}
\maketitle

\begin{abstract}
The quenching of star formation in galaxies is an important aspect of galaxy evolution, but the physical mechanisms that drive it are still not understood. Measuring the spatial distribution of quenching can help determine these mechanisms.
We present the star-formation histories (SFHs) and stellar metallicity evolution of rapidly quenched regions in 86 local post-starburst (PSB) galaxies from the MaNGA integral field survey, obtained through Bayesian full spectral fitting of their rest-frame optical spectra. We found that regardless of spatial location, the PSB regions have similar past SFHs and chemical evolution, once radial metallicity gradients are accounted for. 
This suggests that all PSB regions are regulated by a common set of local scale processes in the interstellar medium, regardless of the broader triggering mechanism. 
We show that the centres of galaxies with outer PSB regions are also quenching. The central specific star-formation rate (sSFR) has declined by $\sim1.0\;$dex on average during the last 2 Gyr, a significantly steeper decline than main sequence galaxies over the same period ($\approx0.2\;$dex). 
This central quenching can be either synchronous, outside-in or inside-out, and slower or as fast as the outer regions, highlighting the diversity of quenching pathways for local galaxies. 
Our results imply a primary quenching mechanism that is both catastrophic and global in rapidly halting star formation in local galaxies. 
We suggest the predominant cause is galaxy mergers or interactions, with large scale feedback from a starburst or a central supermassive black hole playing a lesser role. 

\end{abstract}




\section{Introduction} \label{sec:intro}
Large scale galaxy surveys of the past decades have shown that red quiescent galaxies have grown in both number and total stellar mass since before cosmic noon \citep{ilbert2013,muzzin2013}, thus requiring blue star-forming galaxies to eventually shut down their star-formation (quenching) and transition into red quiescent galaxies. Numerous mechanisms that can quench a galaxy have been proposed in the literature, but a definitive consensus on their relative importance has yet to be reached \citep[see][and references therein]{man2018}. In addition, the relative importance is likely dependent on galaxy properties such as redshift, stellar mass (density) and environment \citep{veilleux2005,wetzel2013,owen2019,mao2022}. 

One useful approach to probe the physical quenching mechanisms is to study galaxies that recently underwent rapid quenching following a duration of increased star formation, known as post-starburst (PSB) galaxies \citep[e.g.][]{wild2009,rowlands2018a,li2019,zheng2020,paper1}. 
Historically, PSBs were first referred to as ``E+A'' or ``K+A'' galaxies, and were identified through the strong Balmer absorption lines and moderately blue $B-V$ colour in their optical spectra \citep{dressler1983}. 
These spectral features are a result of these galaxies' distinct star-formation history (SFH), where the recent rapid quenching leads to an abundance of A and F type stars but a lack of the shorter lived O and B stars. 

PSBs represent only $\sim1$\% of the galaxy population at $z<1$ \citep{pawlik2016,rowlands2018a}. 
However, due to the short visibility duration of the PSB spectral features after a galaxy's recent quenching, both observational and simulation approaches have shown that the PSB phase could be a significant quenching pathway for galaxies at all redshifts \citep{wild2009,wild2016, rowlands2018a,whitaker2012,belli2019,taylor2023,park2023}. 
Very recent observations from the James Webb Space Telescope (JWST) of early massive quiescent galaxies have also found many galaxies to have PSB-like SFHs at cosmic noon \citep{slob2024,park2024} and $z\geq3$ \citep{carnall2023a,carnall2023b,carnall2024,kimmig2025,nanayakkara2024}. 
Therefore, studying PSB galaxies is key to understanding the physical processes driving rapid quenching at both high and low redshifts.

Various physical mechanisms have been proposed to cause the rapid quenching in PSBs \citep[for a recent review, see][]{french2021_review}. In dense environments such as galaxy clusters, mechanisms include the repeated interactions with neighbouring galaxies \citep[``harassment'',][]{moore1996,moore1998} and the rapid removal of the in-falling galaxy's gas reservoir \citep[ram-pressure stripping,][]{poggianti1999,werle2022}. In lower density field and group environments, mechanisms that lead to a removal of the galaxy's cold interstellar medium gas either through ejection or (near) complete consumption through star formation have been widely discussed, which includes AGN feedback \citep{pontzen2017,davis2019,zheng2020,wild2020} and stellar feedback or consumption \citep{wild2009}. Mergers are thought to play a role in triggering the initial starburst and the subsequent quenching of PSBs \citep{bekki2005,wild2009,snyder2011,pawlik2019,davis2019,zheng2020,ellison2022,ellison2024}, supported by the observation of disturbed morphologies and tidal features in many PSBs \citep{zabludoff1996,chang2001,pawlik2016,pawlik2018,wilkinson2022,ellison2024}.

It remains challenging to deduce the relative importance of different quenching mechanisms in forming PSBs. This task have been further complicated by recent discoveries from resolved observations of local PSBs. 
In the past, PSBs have only been identified and classified by considering the galaxy as a whole or central regions limited by the spectroscopic fibre diameter. Recent spatially-resolved Integral Field Spectroscopy such as MaNGA \citep{MANGA} have revealed the complex structures within PSBs. \cite{chen2019} analysed galaxy spectra from individual spaxels (spatial-pixels) from the MaNGA survey and found 360 galaxies that contain regions with PSB signatures. Out of those, they define 31 of them as galaxies with central contiguous post-starburst regions in the galaxy (CPSB, closest to the historical PSBs observed in slit or single fibre spectroscopy), 37 with a central region that has few PSB spaxels surrounded by a ring of PSB spaxels (RPSB) and the remaining 292 with irregular post-starburst regions (IPSB). The number of galaxies with IPSB regions suggests that the rapid quenching required to produce PSB signatures need not be a galaxy-wide process, but can happen in localized regions. 

A new question thus arises: do PSBs with different spatial distributions of PSB regions have different origins with distinct quenching mechanisms? Alternatively, are they driven by the same quenching mechanisms, but are in different stages of the same process? If the former is true, differences in the PSB types' bulk properties (e.g. stellar mass, metallicity) and environment can provide clues on why the galaxies quenched differently. If the latter is true, the common quenching mechanism must create the evolution observed in the PSB regions' spatial distribution. In either case, the answering of the question will provide new constraints on the physical mechanism(s) that quench local PSBs. 

From measuring the radial gradients of spectral indices $\mathrm{D_n} 4000$, $\mathrm{H}\delta_\mathrm{A}$ and $\mathrm{W(H}\alpha)$ as proxies for the stellar age gradient, \cite{chen2019} and later \cite{cheng2024} found that both CPSBs and RPSBs in MaNGA generally have younger stellar populations in their central regions, which is consistent with the two PSB types as two stages of the same outside-in quenching process where RPSBs could evolve into CPSBs. However, \cite{chen2019} also found non-matching radial profiles in mass-weighted age and stellar $v/\sigma$ between CPSBs and RPSBs, thus concluding that CPSBs and RPSBs might have different origins. Comparing the molecular gas mass fractions and kinematics of 9 CPSBs to 4 RPSBs drawn from the same MaNGA sample, \cite{otter2022} observed that CPSBs tend to have more suppressed SFRs for their gas mass fractions, and have more misaligned gas and stellar rotation than RPSBs. They concluded that both PSB types could be remnants of recent mergers, while RPSBs could result from a less disruptive merger or a more rapid resettling of gas, but a large sample is needed.

The conclusions from \cite{chen2019} and \cite{cheng2024} partially relied on using spectral indices as proxies for stellar age. These indices, while proven to be powerful tools in inferring broad shapes of the recent SFHs \citep[e.g.][]{kauffmann2003a}, provide insufficient temporal resolution to confidently determine the evolutionary connection between CPSBs and RPSBs. Therefore, in this study we obtain accurate SFHs of both CPSB and RPSBs from the local Universe through exploiting the full optical spectra using Bayesian spectral fitting methods. Combined with measurements of the chemical evolution of both PSBs types, we aim to deduce the evolutionary relationship between CPSBs and RPSBs, and their likely origins, which has implications on the physical mechanisms that quench star formation.

The structure of this article is as follows. 
In Section \ref{sec:CPSBvsRPSB_data}, we describe the data and sample selection from the MaNGA survey. 
In Section \ref{sec:fitting}, we document the process of optical continuum spectral fitting for the PSB regions in CPSBs and RPSBs, and the central regions of RPSBs, from which we measure their SFH and stellar metallicity evolution. 
The results are presented in Section \ref{sec:results}, and their implications for the evolution of PSBs and their physical quenching mechanisms are discussed in Section \ref{sec:CPSBvsRPSB_discussion}. 
Where necessary, we assume a cosmology with $\Omega_M=0.3$, $\Omega_\Lambda=0.7$ and $h = 0.7$.
All magnitudes are in the AB system \citep{oke1983}.
We assume a  \citet{kroupa2001} stellar initial mass function (IMF), and take solar metallicity $Z_\odot=0.0142$ \citep{asplund2009}. We re-scale all metallicity measurements quoted from the literature to this solar metallicity for direct comparison.

\section{Data} \label{sec:CPSBvsRPSB_data}
\begin{figure*}
	\includegraphics[width=\textwidth]{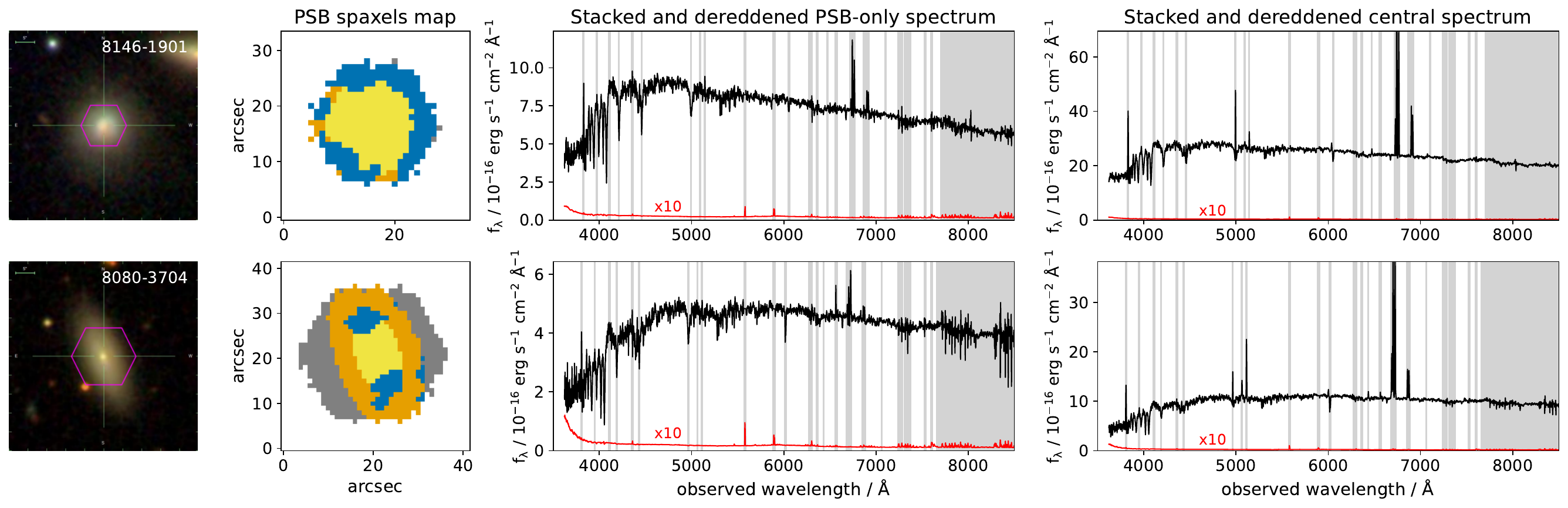}
    \caption{Two typical RPSBs from our sample. The left-most panels show the SDSS 3-colour image with the galaxy's Plate-IFU marked on the top right corner. The MaNGA field of view is marked as the pink hexagon. The middle-left panels show the spaxel selection, displaying regions with no/faulty observations (transparent), with median spectral $\mathrm{SNR}<8$ too low to be classified (grey), classified as PSB (blue), classified as non-PSB (orange) in Section \ref{sec:CPSBvsRPSB_data}; and classified as the central region (yellow) in Section \ref{sec:data_central}. The middle-right and right-most panels show the stacked observed-frame spectrum of the PSB classified spaxels and the central region, respectively. In both columns, we show the stacked spectrum (black), the stacked $1\sigma$ observational uncertainty (red, multiplied by $10\times$ to make visible) and spectral ranges masked during the fitting process (grey bands), including major nebular emission lines, skyline residuals and Balmer infilling. The top RPSB is likely face-on, so its PSB regions show a clear ring shape. The bottom RPSB is likely edge-on, so its PSB regions are in two distinct lobes.}
    \label{fig:RPSB_masking}
\end{figure*}

We select CPSBs and RPSBs from the integral field spectrograph MaNGA survey \citep{MANGA}. MaNGA is a part of the fourth-generation Sloan Digital Sky Survey \citep[SDSS-IV,][]{SDSS_IV} and observed $\approx10000$ $M_* > 10^9 M_\odot$ galaxies (11273 datacubes) in the local $z<0.2$ neighbourhood from 2014 to 2020. 
To select PSBs, we broadly follow the procedure outlined in Section 2 of \cite{paper1}. We first apply a redshift cut of $z<0.06$ on all MaNGA galaxies in the data release DR17 \citep{SDSS_DR17}, in favour of local PSBs with good spatial resolution. After removing spaxels marked with \texttt{NOCOV}, \texttt{LOWCOV} or \texttt{DONOTUSE} flags by the data reduction pipeline \citep{manga_drp,manga_drp_new}, we identify PSB spaxels with a similar selection to that in \cite{chen2019}, specifically requiring the spaxels' median spectral $\mathrm{SNR} > 8$ per pixel, strength of the $\mathrm{H\delta}$ 
Balmer absorption line after accounting for emission infilling $\mathrm{H\delta_A}>3\text{\normalfont\AA}$ \citep{worthey1997}, equivalent width of the H$\alpha$ nebular emission line after accounting for underlying absorption $\mathrm{W(H\alpha)}<10\text{\normalfont\AA}$\footnote{$\mathrm{W(H\alpha)}$ follows the passband definitions in \cite{manga_dap}, which is a modified version of passbands in \cite{yan2006}.}, and $\log \mathrm{W(H\alpha)}< 0.23\times \mathrm{H\delta_A} - 0.46$. Both $\mathrm{H\delta_A}$ and $\mathrm{W(H\alpha)}$ are obtained from the MaNGA data analysis pipeline \citep[DAP][]{manga_dap,manga_dap2}. 

\begin{figure}
    \centering
    \includegraphics[width=\columnwidth]{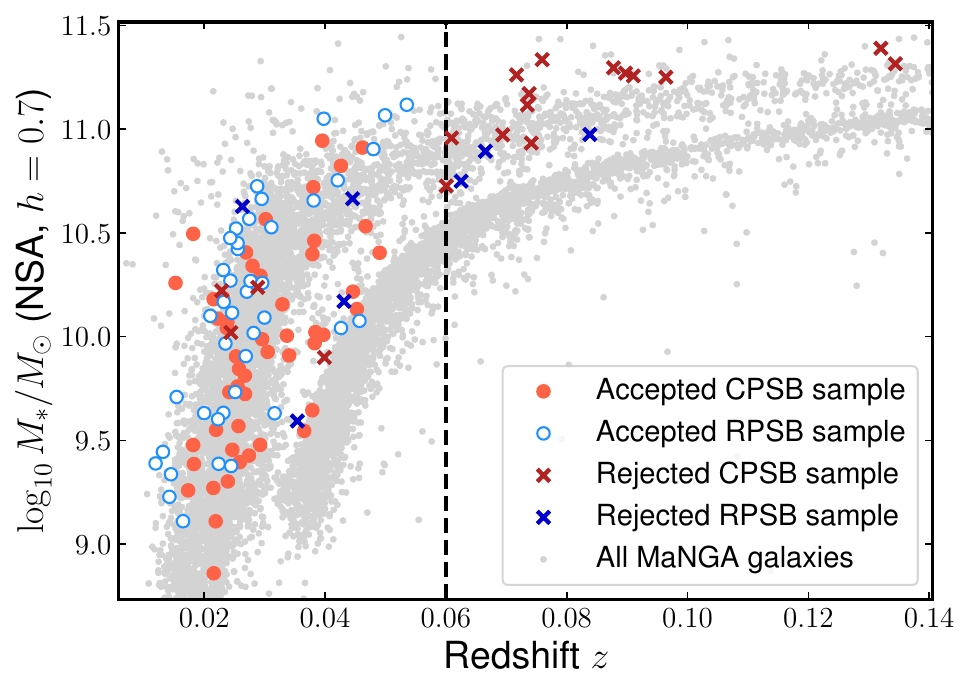}
    \caption{The redshift and stellar mass distributions of the CPSB (red dots or crosses) and RPSB (blue dots or crosses) galaxies with global PSB spaxel fraction $>0.05$. For both PSB types, dots mark candidates that form this study's final sample, while crosses mark candidates removed from the sample. To the right of the vertical dashed line, which marks $z=0.06$, the candidates were removed due to the redshift cut. To the left of the vertical dashed line, the candidates were manually removed due to issues related to their observational quality or spectral properties (see main text). The distribution of all MaNGA galaxies is shown in the background for reference. The stellar masses from the NSA catalogue have been adjusted to $h = 0.7$.}
    \label{fig:data_redshift_vs_Mstar}
\end{figure}

To split the sample into CPSBs and RPSBs, we sliced the galaxies into 3 elliptical annuli with $0<R/R_e<0.5$, $0.5<R/R_e<1$ and $1<R/R_e<1.5$, where the radial distance $R$ is obtained from the MaNGA data analysis pipeline, and was derived from the ellipticity and position angle of the galaxy's $r$-band surface-brightness maps \citep{manga_dap}. $R_e$ is the $r$-band elliptical-Petrosian effective radius. The CPSB galaxy sample is selected by requiring $>50$\% of the inner annulus spaxels classifiable, $>50$\% of these spaxels to be classified as a PSB, and a global PSB spaxel fraction of $>0.05$ among all classifiable spaxels, which yields 54 candidates. The RPSB galaxy sample is selected by the following three criteria:
\begin{compactlists}
    \item[1.] Has a global PSB spaxel fraction of $>0.05$ among all classifiable spaxels
    \item[2.] Has $>50$\% of the inner annulus spaxels classifiable, and $<40$\% of these spaxels to be classified as a PSB
    \item[3.] Fulfils either one of the following:
    \begin{compactlists}
    \item[a.] Has $>40$\% of the middle annulus spaxels classifiable, and $>40$\% of these spaxels to be classified as a PSB
    \item[b.] Has $>40$\% of the ouer annulus spaxels classifiable, and $>30$\% of these spaxels to be classified as a PSB
    \end{compactlists}
\end{compactlists}
This yields 45 RPSB candidates, two of which are shown in Figure \ref{fig:RPSB_masking} (see the central left panels for a visualization of selecting RPSB regions). We plot the redshift and stellar mass distributions of these PSB candidates in Figure \ref{fig:data_redshift_vs_Mstar}. The stellar masses are adopted from the NASA-Sloan Atlas \citep[\texttt{NSA\_ELPETRO\_MASS}, a K-correction fit to elliptical Petrosian fluxes, see][]{blanton2011}, adjusted to $h = 0.7$.

We removed candidates with undesirable MaNGA observations (e.g. mismatched redshift and obvious foreground stars, removed 4: 7972-12702, 8248-6104, 8601-12703, 8935-12703), active galactic nuclei (AGN) broad emission (removed 2: 8723-6104, 11004-6104) and datacubes flagged as \texttt{BADFLUX} by the MaNGA DRP (removed 1: 8944-1902). The final sample of 50 CPSBs have been listed in Table 1 of \cite{paper1}. RPSB 11952-3703 has two clearly defined lobes, one of which appears to have a significantly higher redshift than the other, likely due to a background galaxy. Therefore, we remove this galaxy from our sample, yielding a final sample of 41 RPSBs. The RPSBs span a total stellar mass range of $9.11<\log_{10}M_*/M_\odot<11.12$, slightly higher than our CPSB sample ($8.86<\log_{10}M_*/M_\odot<10.94$, see Figure \ref{fig:data_redshift_vs_Mstar}). The RPSBs' stellar mass and other properties are listed in Table \ref{tab:RPSB_targets}.

To ensure spaxel quality, we remove spaxels marked with quality flags \texttt{DEADFIBER} or \texttt{FORESTAR} in MaNGA's H$\alpha$ emission line maps. Next, we form a stacked PSB-only spectrum for each galaxy (both CPSBs and RPSBs) and a stacked spectrum from the central region for each RPSB. The PSB-only spectra are obtained by summing the spectra of all spaxels classified as PSB from our spaxel selection, while the stacking of the central regions is detailed in Section \ref{sec:data_central}. In both stacks, spectra are summed unweighted, while uncertainties are summed in quadrature. After correcting for Milky Way dust reddening assuming the Galactic extinction law presented by \cite{milky_way_reddening}, we mask major nebular emission lines, residuals of strong skylines (central flux $>5 \times 10^{-16} \ \mathrm{erg\ s^{-1}\ cm^{-2}\ \text{\normalfont\AA}^{-1}}$ in \citealp{skylines}) and Balmer infilling. Following \cite{paper1}, to restrict the wavelength range of spectra fitted to fully within the MILES stellar library's range (\citealt{MILES}, see also \citealt{pawlik2018}), we limit all spectra to rest frame $\lambda<7500\text{\normalfont\AA}$.
This process yields high SNR stacked spectra that represent only the PSB regions in each galaxy or the central regions in each RPSB. 
The middle right and right-most columns in Figure \ref{fig:RPSB_masking} show the PSB-only and central stacked spectra of two example RPSBs, respectively.

\begin{table*}
    \centering
    \caption{List of 41 studied ringed post-starburst (RPSB) galaxies and their properties: 
    (1) MaNGA Plate-IFU identifier;
    (2) MaNGA identifier;
    (3) R.A. (J2000);
    (4) Declination (J2000);
    (5) Redshift;
    (6) $\log_{10}$ total stellar mass fitted from K-corrected elliptical Petrosian photometric fluxes in GALEX/SDSS \textit{FNugriz} bands from the NSA catalogue, adjusted for $h=0.7$;
    (7) Number fraction of classified PSB spaxels among all spaxels not marked with the \texttt{NOCOV} or \texttt{LOWCOV} flags in MaNGA datacubes;
    (8) Final number of stacked PSB spaxels, after excluding spaxels marked with \texttt{DEADFIBER} or \texttt{FORESTAR};
    (9) Mean SNR of the stacked PSB spectrum over the full MaNGA wavelength range; 
    (10) Final number of stacked central spaxels, after excluding spaxels marked with \texttt{DEADFIBER} or \texttt{FORESTAR}; 
    (11) Mean SNR of the stacked central spectrum over the full MaNGA wavelength range.
    The full table is available as supplementary online material.}
    \begin{tabular}{p{1.4cm}p{1.1cm}p{1.2cm}p{1.2cm}p{1.1cm}p{1.1cm}p{1.0cm}p{1.3cm}p{1.3cm}p{1.3cm}p{1.3cm}}
    \\\hline
    Plate-IFU \newline(1) & MaNGA ID \newline(2) & RA \newline(degrees) \newline(3) & Dec. \newline(degrees) \newline(4) & Redshift \newline(5) & $\log_{10}M_*$ ($h=0.7$) \newline(6) & PSB spaxel fraction \newline(7) & Number of stacked PSB spaxels \newline(8) & PSB stacked mean SNR \newline(9) & Number of stacked central spaxels \newline(10) & Central stacked mean SNR \newline(11) \\ \hline
    7990-12705  & 1-24476   & 264.68032 & 57.23921  & 0.0295    & 10.66     & 0.17      & 447       & 333.7     & 162       & 453.6              \\
    8080-3704   & 1-38041   & 49.45745  & -0.55466  & 0.021     & 10.1      & 0.11      & 86        & 238.0     & 101       & 421.7              \\
    8083-12703  & 1-38168   & 49.92934  & 0.56548   & 0.0243    & 10.27     & 0.13      & 347       & 292.3     & 423       & 578.9              \\
    8146-1901   & 1-201355  & 117.05387 & 28.22509  & 0.0271    & 10.22     & 0.46      & 176       & 308.8     & 181       & 753.0              \\
    8152-3703   & 1-232087  & 142.72474 & 35.43978  & 0.0277    & 10.27     & 0.2       & 157       & 192.4     & 236       & 468.1              \\
    \multicolumn{11}{c}{$\ldots$} \\
    \hline
    \end{tabular}
    \label{tab:RPSB_targets}
\end{table*}

\subsection{Stacking the central regions of RPSBs} \label{sec:data_central}
For each RPSB, we sum the central region surrounded by the PSB regions into a stacked spectrum. We define the selection criteria of this region as spaxels with median spectral $\mathrm{SNR} > 8$ per pixel, not already classified as PSB, and has in-plane elliptical polar radial distances from the galaxy centre lower than the median value of the PSB spaxels.

The central region and the corresponding stacked spectra of the two example RPSBs are shown in Figure \ref{fig:RPSB_masking}. Comparing the stacked PSB-only and central spectra, the central regions of these two RPSBs have much stronger H$\alpha$ and H$\beta$ emission and slightly weaker Balmer absorption strengths, indicating that they likely have stronger ongoing star formation than the PSB regions. 

It is important to be aware of AGN contamination in our stacked spectra, particularly in the central regions. This is because emission from the AGN (mainly through the blue continuum) may change the normalization and shape of the galaxy spectrum observed \citep[e.g.][]{francis1991,kauffmann2003b,davies2016}, leading to incorrect measurements of galaxy properties using our models that do not account for AGN contribution. 

To obtain a list of the RPSBs with possible AGN contamination, we combined the MaNGA AGN catalogues from \cite{comerford2020}, \cite{comerford2024} and \cite{alban2023}, along with constructing WHAN diagrams to classify the ambiguous galaxies \citep{cid-fernandes2011}. 
We select Seyfert RPSBs from all galaxies classified as AGN candidates in the \cite{comerford2024} catalogue or as Seyfert in the \cite{alban2023} catalogue, yielding 5/41 galaxies. We also select RPSBs with possible LIER/LINER contamination from galaxies classified as LINER, composite or ambiguous in the \cite{alban2023} catalogue, and requiring $>50\%$ of the central $R<0.5R_e$ spaxels occupying the strong or weak AGN regions of the WHAN diagram, yielding 12/41 galaxies.

Although we labelled a high fraction of RPSBs (17/41) where their central regions might be contaminated by AGNs, we found no observable difference in our results between RPSBs without AGNs and with a possible AGN. Galaxies with central LIER/LINER signatures make up most of the sample we marked as possible AGNs. Our results would not be affected by LIER/LINER contamination as we mask all major non-Hydrogen emission lines and the H$\alpha$-[NII] complex during spectral fitting. Therefore, we do not expect AGN contamination to have a significant effect on our fitted SFHs.

\section{Optical continuum spectral fitting} \label{sec:fitting}
To obtain the SFH, stellar metallicity evolution and dust attenuation of the regions we selected from the MaNGA PSBs, we perform Bayesian full spectral fitting on the high SNR stacked spectra described in Section \ref{sec:CPSBvsRPSB_data}. The fitting methods for all PSB regions and the central regions of RPSBs differ slightly. Therefore, here we first describe the components common to both and in the following sections the components unique to each region is discussed.

In all spectral fitting in this study, we use the Bayesian spectral energy density (SED) fitting code \textsc{Bagpipes} \citep{bagpipes2018,bagpipes2019}, which is based on the \cite{bruzual2003} stellar population synthesis models (2016 version), and assume the initial mass function from \cite{kroupa2001}. We apply the two-component dust attenuation law from \cite{wild2007} and \cite{dacunha2008}, with a fixed power-law exponent $n=0.7$ for the interstellar medium (ISM). The dust law asserts that stars younger than $10\,$Myr have a steeper power-law exponent $n=1.3$ and are more attenuated than older stars by a factor $\eta$ \citep[$=1/\mu$ in][]{wild2007,dacunha2008}, as they are assumed to be surrounded by their birth clouds. The model priors are listed in the top section of Table \ref{tab:priors}.

\begin{table*}
\centering
	\caption{Model priors used for fitting SEDs. The model components in the top section are used in all fits. The components specific to the PSB regions of both CPSBs and RPSBs are listed in the middle section, while the components specific to the central regions of RPSBs are listed in the bottom section. The parameter symbols are described in Section \ref{sec:fitting}, or otherwise have their usual meanings. Some parameters have prior shape $\log_{10}$ uniform, which indicates a flat prior in uniform space $\log(X) \sim U(\log(min), \log(max))$. Redshift is given a uniform prior ranging from 80\% to 120\% of the target's MaNGA redshift ($z$). Note that $\sigma_{\mathrm{disp}}$ is not the intrinsic velocity dispersion of the galaxy, as it does not account for the finite resolution of the spectral templates or observational data.}
	\label{tab:priors}
    \begin{tabular}{llllll}
        \hline
        Usage & Type & Parameter & Form & Min & Max \\ 
        \hline
        All & Dust & $A_V$ / mag & Uniform & 0 & 3 \\
         &  & birthcloud factor $\eta$ & Uniform & 1 & 5 \\
         &  & $t_{\rm{birth cloud}}$ / Gyr & Fixed = 0.01 & - & - \\
         & GP noise & uncorrelated amplitude $s$ & $\log_{10}$ Uniform & 0.1 & 10 \\
         &  & correlated amplitude $\sigma$ & $\log_{10}$ Uniform & $10^{-4}$ & 1 \\
         &  & period/length scale $\rho$ & $\log_{10}$ Uniform & 0.04 & 1.0 \\
         &  & dampening quality factor $Q$ & Fixed = 0.49 & - & - \\ 
         & Miscellaneous & redshift & Uniform & 0.8 $z$ & 1.2 $z$ \\
         &  & $\sigma_{\rm{disp}}$ / km/s & $\log_{10}$ Uniform & 40 & 4000 \\
         \hline
        PSB regions & SFH & $\log_{10}(M_*/M_\odot)$ & Uniform & 6 & 13 \\
         &  & $t_{\rm{form}}$ / Gyr & Uniform & 4 & 14 \\
         &  & $\tau_e$ / Gyr & Uniform & 0.3 & 10 \\
         &  & $t_{\rm{burst}}$ / Gyr & Uniform & 0 & 4 \\
         &  & $\alpha_\mathrm{PSB}$ & $\log_{10}$ Uniform & 0.01 & 1000 \\
         &  & $\beta_\mathrm{PSB}$ & Fixed = 250 & - & - \\
         &  & $f_{\rm{burst}}$ & Uniform & 0 & 1 \\
         & Metallicity & $Z_{\rm{old}}/Z_\odot$ & $\log_{10}$ Uniform & 0.014 & 3.52 \\
         &  & $Z_{\rm{burst}}/Z_\odot$ & $\log_{10}$ Uniform & 0.014 & 3.52 \\ 
         \hline
        Central regions & SFH & $\log_{10}(M_*/M_\odot)$ & Uniform & 6 & 13 \\
         &  & falling slope $\alpha_c$ & $\log_{10}$ Uniform & 0.01 & 250 \\
         &  & rising slope $\beta_c$ & $\log_{10}$ Uniform & 0.01 & 250 \\
         &  & turnover time $\tau$ / Gyr & Uniform & 0 & 14 \\
         & Metallicity & $Z_*/Z_\odot$ & $\log_{10}$ Uniform & 0.014 & 3.52 \\
         & Nebular & ionization parameter $\log_{10}U$ & Fixed = 3 & - & - \\
         \hline
    \end{tabular}
\end{table*}

\subsection{The PSB regions of CPSBs and RPSBs} \label{sec:RPSB_fitting}
The spectral fitting on the PSB-only stacked spectra of both CPSBs and RPSBs follows the methods detailed in \cite{paper1}. Here, a brief summary is provided. 

In addition to the model components listed above, we adopt the two-component parametric SFH model for PSBs from \cite{wild2020}:
\begin{equation}\label{eq:psb2}
    \mathrm{SFR}(t) \propto \frac{1-f_{\mathrm{burst}}}{\int \psi_e \mathrm{d}t} \times \psi_e(t)\Big|_{t_{\mathrm{form}}>t>t_{\mathrm{burst}}} 
    + \frac{f_{\mathrm{burst}}}{\int \psi_{\mathrm{burst}} \mathrm{d}t} \times \psi_{\mathrm{burst}}(t) \; .
\end{equation}
The SFH model is a sum of an older, exponential decay component that models the existing stellar populations pre-starburst ($\psi_e$), and a younger, double power-law component that models the recent starburst and its quenching ($\psi_{\mathrm{burst}}$). Both components are functions of lookback time $t$. $t_{\mathrm{form}}$ is the lookback time when the older population began to form; $t_{\mathrm{burst}}$ is the time since the peak of the starburst; the fraction $f_{\mathrm{burst}}$ is the proportion of mass formed during the starburst. The two components are given by:
\begin{align}
\label{eq:exp}
    \psi_e(t') &= \exp^{\frac{-t'}{\tau_e}} \\
\label{eq:dpl}
    \psi_{\mathrm{burst}}(t') &= \Big[\big(\frac{t'}{t'_{\mathrm{burst}}}\big)^{\alpha_\mathrm{PSB}} 
    + \big(\frac{t'}{t'_{\mathrm{burst}}}\big)^{-\beta_\mathrm{PSB}}\Big]^{-1}\;,
\end{align}
where $t'$ is the age of the Universe, $t'_{\mathrm{burst}}$ is the age of the Universe at the peak of the starburst, $\tau_e$ is the e-folding timescale of the exponential component, while $\alpha_\mathrm{PSB}$ and $\beta_\mathrm{PSB}$ control the declining and increasing timescales of the burst respectively, with larger values corresponding to steeper slopes. During fitting, we set the total stellar mass formed ($\log_{10}(M_*/M_\odot)$), $t_{\mathrm{form}}$, $\tau_e$, $t_{\mathrm{burst}}$, $f_{\mathrm{burst}}$, and $\alpha_\mathrm{PSB}$ as free parameters, while we fix $\beta_\mathrm{PSB}=250$ following \cite{wild2020} and \cite{paper1}.

To model the change in metallicity during the recent starburst, we assume a two-step model for the evolution in stellar metallicity of newly formed stars that transitions at lookback time $t=t_\mathrm{burst}$, and no scatter in coeval stars. This metallicity model has been shown in \cite{paper1} to recover unbiased estimates of both the pre- and post-burst metallicity levels and improve SFH measurements. To account for correlated observational uncertainties and model-data mismatch stemming from observational and callibrational issues and imperfect stellar templates, we employ a Gaussian Process noise model implemented through the \texttt{celerite2} python package \citep{celerite,celerite2} as an additive correctional term across the observed wavelength range. The sampling of the posterior surface is done using the \textsc{MultiNest} nested sampling algorithm \citep{multinest} and its python interface \citep{pymultinest}. The detailed model priors can be found in the middle section of Table \ref{tab:priors}.

We fit the PSB regions of all 41 RPSBs, and use the results from \cite{paper1} for the CPSBs. 
Unlike the CPSB sample, which had 5/50 ($10\%$) that resulted in fitted GP noise components showing obvious trends across the fitted spectral range and/or with amplitudes much larger than the scaled observational uncertainty, none of the RPSB sample had problematic GP noise fit results. 
All 41 were found to have clear PSB SFHs, where rapid quenching follows an episode of strong starburst.

\subsection{The central regions of RPSBs} \label{sec:central_fitting}
The main goal from fitting the central regions' stacked spectra is to obtain bulk properties, such as the current SFR, sSFR and mass-weighted age, instead of its detailed SFH. These regions could exhibit a wide range of SFH shapes, from a declining SFH similar to a quiescent galaxy to a rising SFH with active current star formation. We therefore use the parametric double power law SFH model, which has been shown to be a robust model for a wide range of SFH shapes including rising and declining SFHs \citep{carnall2019}. It has the form:
\begin{equation}
    \mathrm{SFR}(t') = \Big[\big(\frac{t'}{\tau}\big)^{\alpha_c} 
    + \big(\frac{t'}{\tau}\big)^{-\beta_c}\Big]^{-1}\;,
\end{equation}
where $t'$ is the age of the Universe, $\tau$ is the turnover age of the Universe, $\alpha_c$ is the coefficient for the falling slope, and $\beta_c$ is the coefficient for the rising slope.

To maintain consistency with the outer PSB regions, in Appendix \ref{apx:central_sfh} we investigated also using the \cite{wild2020} SFH model for the central regions. However, accurate recovery of the SFH suffered due to the assumed fixed rising slope of the recent starburst ($\beta$ parameter). We also repeated our central region SED fits with the non-parametric ``continuity'' SFH model shown by \cite{leja2019} to be a robust SFH model for a wide range of true SFH shapes. For most central regions, this yields no significant change in the fitted SFHs. Considering that the double power law model requires less computation per fit and provides a more intuitive comparison to the outer PSB regions' SFHs (e.g. comparing when the recent quenching began), we favour the former model to fit the central regions.

For the central regions of RPSBs, for simplicity, we assume a constant metallicity evolution model with no scatter in coeval stars. Preliminary testing with the GP noise model used in the PSB regions yielded problematic fitted GP noise correction spectra. Most exhibited a continuous slope spanning the whole wavelength range, or had corrections with magnitudes significantly higher than the observational uncertainty at that wavelength. Therefore, we update our likelihood function with additional constraints while fitting these central regions to limit the magnitude of this GP noise correction. For reference, the original log-likelihood function implemented in \textsc{Bagpipes} as detailed in \cite{bagpipes2019} is
\begin{equation}\label{eq:GP_likelihood}
    \ln(\mathcal{L}_{\mathrm{spec}}) = D - \ln(|\mathbf{C}(\Phi)|) - \Delta^\intercal \mathbf{C}(\Phi)^{-1}\Delta \; ,
\end{equation}
where $D$ is a constant correlated with the determinant of the covariance matrix $\mathbf{C}(\Phi)$, $\Phi$ are the parameters of the GP noise component, and $\Delta = y_j - m_j(\Theta)$ is the vector of difference between the observed ($y_j$, $j$ runs over the number of pixels in the spectrum, $N_\mathrm{pix}$) and the model-predicted SED fluxes with no GP noise contributions ($m_j(\Theta)$, where $\Theta$ are the physical model parameters).

Here we modify the log-likelihood function to become:
\begin{equation}\label{eq:clamped_GP_likelihood}
    \ln(\mathcal{L}_{\mathrm{spec}}) = (D - \ln(|\mathbf{C}(\Phi)|) - \Delta^\intercal \mathbf{C}(\Phi)^{-1}\Delta)\Big[\max\Big(1, \frac{\chi^2_\nu}{\gamma}\Big)\Big]
\end{equation}
where $\gamma$ is a constant and $\chi^2_\nu$ is the reduced chi-squared value, given by
\begin{equation} \label{eq:chi_squared_GP}
    \chi^2_\nu = \frac{1}{\nu}\sum^{N_\mathrm{pix}}_j \bigg( \frac{y_j-m_j(\Theta)}{s\cdot\sigma_j} \bigg)^2 \; .
\end{equation}
The degrees of freedom, $\nu = N_\mathrm{pix}-\mu$, equals the number of spectral pixels ($N_\mathrm{Npix}$) minus the number of free parameters ($\mu$). Crucially, the model predicted flux at index $j$ in Equation \ref{eq:chi_squared_GP} ($m_j(\Theta)$) is only from the physical model and excludes contributions from the GP noise. The parameter $s$ is the scaling of the observational uncertainty in Table \ref{tab:priors}. The effect of this modification is such that for sample draws of parameters $\Theta$ that result in model-predicted spectra with reduced chi-squared values greater than $\gamma$ ($\chi^2_\nu>\gamma$), the log-likelihood is inflated by the factor $\chi^2_\nu/\gamma$. While for samples draws that result in model-predicted spectra with $\chi^2_\nu\leq\gamma$, the log-likelihood is unchanged, identical to that from Equation \ref{eq:GP_likelihood}. Since the log-likelihood of any SED fit is typically negative, the modification further down-weights parameter spaces that lead to large differences between the observed and physical model-predicted spectra, thus limiting the fitted GP noise component to be low in magnitude across all wavelengths. 

From comparing the $\chi^2_\nu$ values of fits with unproblematic GP noise corrections to those from fits with problematic GP noise corrections, we found setting $\gamma=3$ yields the best improvements in minimizing the magnitude of the GP noise correction in problematic fits. Moreover, we verified that central PSB regions that originally had unproblematic GP noise corrections remained unproblematic under the modified log-likelihood function, and their posterior distributions remained unaffected. On the other hand, we observed that most central PSB regions with originally problematic GP noise corrections had significantly reduced amplitudes in their new GP noise corrections, and no longer have corrections entirely offset from the zero line for the whole wavelength range or shapes that resemble stellar spectra. With less problematic contributions from the GP noise component, the posterior estimations obtained through the modified log-likelihood function are expected to be more representative of the true stellar and dust properties. 

When fitting the PSB regions that have been selected to have minimal current star formation, we mask all major nebular emission lines and Balmer infilling. However, since some central regions can have on-going star formation, we do not apply masks on the Balmer lines when fitting the central regions. We instead include the nebular component implemented in \textsc{Bagpipes} computed through the \textsc{Cloudy} photo-ionization code \citep{cloudy}. We fix the ionization parameter as $\log_{10}U=-3$.

All other components of the SED fitting model are the same as the PSB fitting method described above, including dust attenuation, redshift and velocity dispersion. The detailed priors are listed in the bottom section of Table \ref{tab:priors}. For all central regions, the sampling of the posterior surface is done using \textsc{nautilus} \citep{nautilus}, a neural-network-boosted nested sampling algorithm, with 800 live points.

Using the framework presented above, we fit the central regions of all 41 RPSBs. 
Despite the improvements made from modifying the log-likelihood function, 4/41 (10\%) resulted in fitted GP noise components that show obvious trends across the fitted spectral range, with amplitudes much larger than the scaled observational uncertainty. 
This indicates that either these galaxies are particularly impacted by data reduction issues or have pathological true SFHs and/or chemical evolution histories that lead to severe model-data mismatch under our simplistic models.
Considering their fitted SFH and galaxy properties are likely biased, we exclude their results from further analysis (Plate-IFU: 8250-12704, 8322-1901, 11749-12704, 11827-1902). We note that the excluded galaxies were also among the top 8 galaxies with the largest residuals if fitted without the GP noise component and with the non-parametric ``continuity'' SFH model, indicating this issue is not related to the usage of the GP noise or the SFH model choice.

\section{Results} \label{sec:results}
\begin{figure}
    \centering
    \includegraphics[width=\columnwidth]{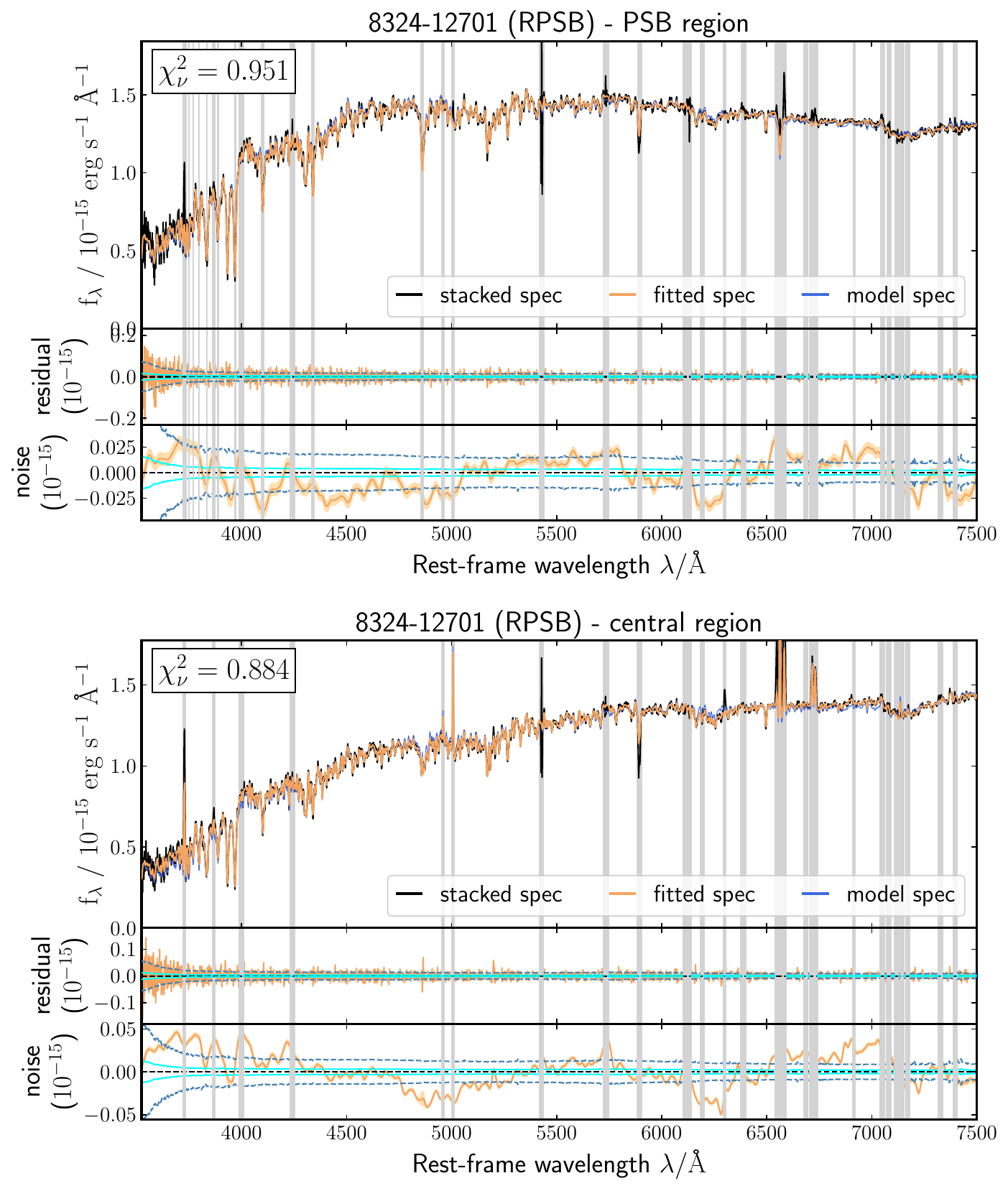}
    \caption{Examples of the fitted stacked MaNGA spectrum of a RPSB (8324-12701), from its outer PSB (\textbf{top panels}) and central (\textbf{bottom panels}) regions. \textbf{Each panel, top}: The observed spectrum created by combining only spaxels from the region (black), the posterior best fit spectrum and its $1\sigma$ region (orange line and orange shaded region), which includes contribution from the physical model (blue) and GP noise. The reduced chi-squared value of the maximum likelihood posterior sample spectrum (including GP noise), $\chi^2_\nu$, is shown. \textbf{Each panel, middle}: The fitting residual (orange), i.e. the observed stacked spectrum (black curve) - posterior best-fit spectrum (orange curve). The input observational uncertainty before (light blue line) and after scaling by the fitted noise scaling factor $s$ (dark blue dashed). An increase of around $\times3-5$ is typically required. \textbf{Each panel, bottom}: The fitted GP noise component and its $1\sigma$ region in orange, with blue curves as above. Fitted GP noise curves with a majority lying within the scaled observational uncertainty (blue dashed) are considered good fits. Note that y-axes have the same units, but the three sub-panels vary in scaling. In all panels, the grey bands indicate masked due to skyline residuals, strong nebular emission lines or Balmer infilling (Balmer lines not masked when fitting central regions).}
    \label{fig:RPSB_fit_example}
\end{figure}

In Figure \ref{fig:RPSB_fit_example}, we show an example fit of the PSB regions of a RPSB. An equivalent example for a CPSB is given in Figure 6 in \cite{paper1}. Thanks to the exceptional SNR of the stacked spectra, we are able to constrain important galaxy properties to within narrow typical uncertainties: $\delta\, \log_{10}M_{*}/M_\odot \approx 0.05\;$dex; $\delta\, t_\mathrm{burst}\approx0.15\;$Gyr; $\delta\, f_\mathrm{burst}\approx0.1$; $\delta\, A_V\approx0.5\;$mag; $\delta\, \log_{10}Z_*/Z_\odot\approx0.15\;$dex. In Table \ref{tab:RPSB_PSB_results}, we report these fitted values and the reduced chi-squared of the maximum likelihood posterior sample spectrum of the PSB regions of our 41 RPSBs. The same results for the PSB regions of the 50 CPSBs are reported in Table 5 of \cite{paper1}. All fitted SFHs and the metallicity evolution of the PSB regions in RPSBs are plotted in Appendix \ref{apx:SFH}. The equivalent CPSB figures can be found in Appendix C of \cite{paper1}.

\begin{table*}
    \centering
    \caption{Posterior estimated properties of the PSB regions of 41 RPSBs from the spectral fitting of stacked MaNGA spaxels. Columns are (1) MaNGA Plate-IFU,
    (2) stellar mass within the stacked PSB spaxels,
    (3) ISM dust attenuation at 5500\AA\ ($V$ band),
    (4) the $\log_{10}$ SFR within the stacked PSB spaxels averaged over the last 100\,Myr,
    (5) time since the peak of the starburst,
    (6) fraction of mass formed during the starburst,
    (7) SFR halving timescale of the starburst,
    (8) stellar metallicity before the burst,
    (9) stellar metallicity during and after the burst,
    (10) change in metallicity,
    and (11) reduced chi-squared value of the maximum likelihood posterior sample spectrum. 
    The full table is available as supplementary online material.}
    \begin{tabular}{lllllllllll}
        \\\hline
        Plate-IFU & $\log_{10}$ & $A_V$/mag & $\log_{10}\mathrm{SFR_{100Myr}}$ / & $t_\mathrm{burst}$/Gyr & \multirow{2}{*}{$f_\mathrm{burst}$ (6)} & $\tau_{1/2}$/Myr & $Z_{\rm{old}}/Z_\odot$ & $Z_{\rm{burst}}/Z_\odot$ & $Z_{\rm{diff}}/Z_\odot$ & $\chi^2_\nu$\\
        (1) & $M_{*,\mathrm{PSB}}/M_\odot$ (2) & (3) & $\mathrm{M_\odot\; yr^{-1}}$ (4) & (5) &  & (7) & (8) & (9) & (10) & (11)\\
        \hline
        7990-12705 & $10.06^{+0.06}_{-0.05}$   & $0.97^{+0.04}_{-0.04}$    & $-4.05^{+2.32}_{-13.44}$  & $1.31^{+0.15}_{-0.10}$    & $0.37^{+0.10}_{-0.10}$    & $123^{+119}_{-87}$        & $0.63^{+0.20}_{-0.17}$    & $0.96^{+0.51}_{-0.30}$    & $0.35^{+0.65}_{-0.50}$    & 0.963  \\
        8080-3704  & $9.31^{+0.05}_{-0.06}$    & $0.77^{+0.04}_{-0.04}$    & $-1.39^{+0.11}_{-0.13}$   & $0.97^{+0.07}_{-0.07}$    & $0.25^{+0.08}_{-0.05}$    & $276^{+36}_{-33}$         & $0.65^{+0.15}_{-0.10}$    & $3.25^{+0.18}_{-0.31}$    & $2.57^{+0.24}_{-0.36}$    & 0.953  \\
        8083-12703 & $9.62^{+0.04}_{-0.06}$    & $0.80^{+0.04}_{-0.03}$    & $-0.33^{+0.05}_{-0.05}$   & $0.88^{+0.14}_{-0.09}$    & $0.21^{+0.07}_{-0.04}$    & $527^{+103}_{-71}$        & $0.43^{+0.04}_{-0.06}$    & $2.59^{+0.31}_{-0.41}$    & $2.16^{+0.33}_{-0.40}$    & 0.946  \\
        8146-1901  & $9.58^{+0.06}_{-0.05}$    & $0.52^{+0.04}_{-0.05}$    & $-0.64^{+0.06}_{-0.06}$   & $0.82^{+0.10}_{-0.06}$    & $0.25^{+0.06}_{-0.05}$    & $308^{+50}_{-36}$         & $0.26^{+0.06}_{-0.03}$    & $2.99^{+0.23}_{-0.26}$    & $2.72^{+0.22}_{-0.27}$    & 0.934  \\
        8152-3703  & $9.58^{+0.04}_{-0.05}$    & $0.50^{+0.05}_{-0.05}$    & $-1.01^{+0.07}_{-0.08}$   & $0.81^{+0.10}_{-0.05}$    & $0.09^{+0.02}_{-0.01}$    & $318^{+53}_{-35}$         & $0.45^{+0.04}_{-0.04}$    & $3.09^{+0.29}_{-0.36}$    & $2.65^{+0.27}_{-0.40}$    & 0.967  \\
        \multicolumn{11}{c}{$\ldots$} \\
        \hline
    \end{tabular}
    \label{tab:RPSB_PSB_results}
\end{table*}

\subsection{CPSBs and RPSBs show similar SFHs in their PSB regions} \label{sec:results1}
\begin{figure*}
    \centering
    \includegraphics[width=0.9\textwidth]{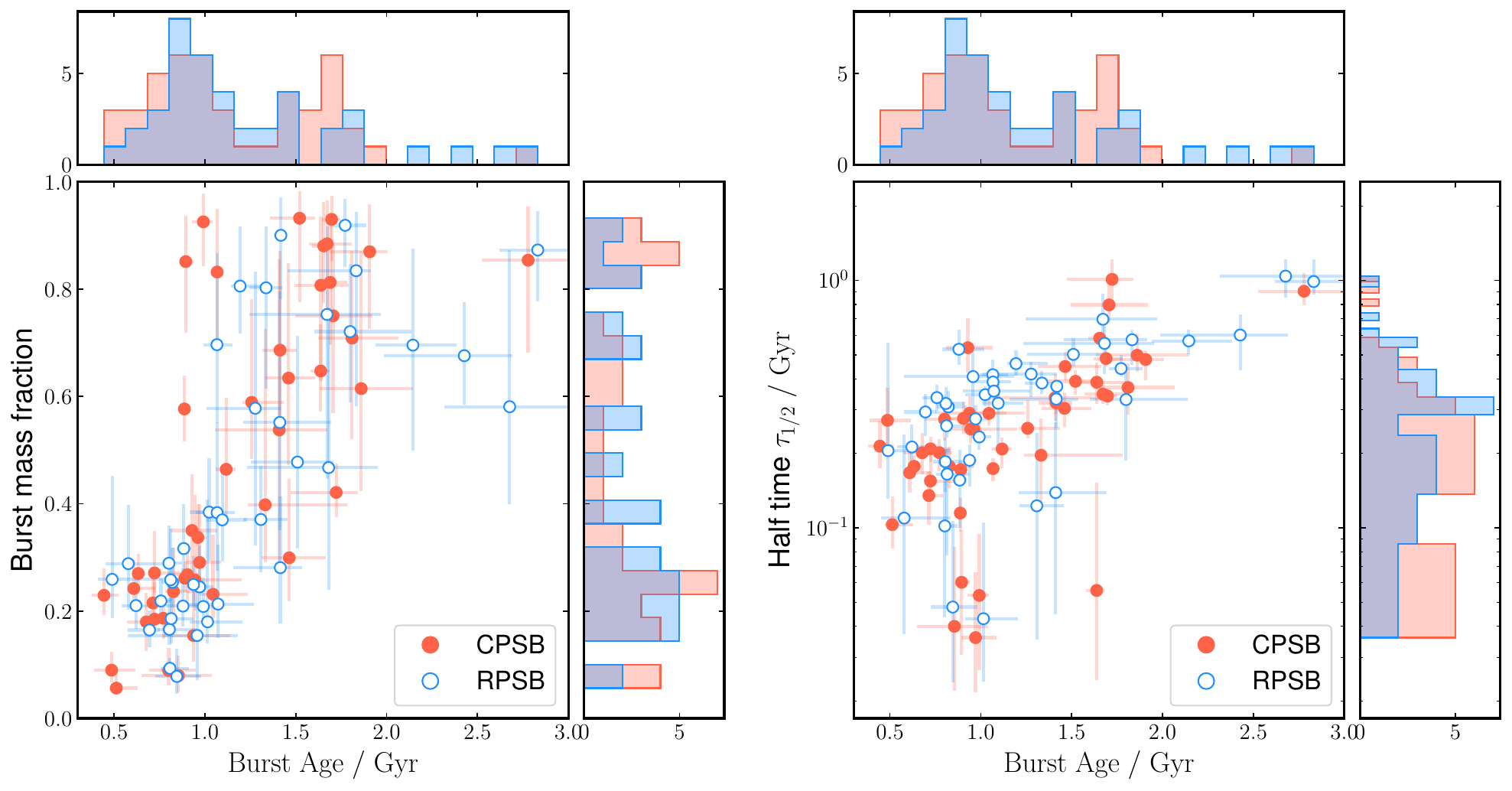}
    \caption{Comparing fitted SFH properties of the PSB regions in CPSBs (red closed dots) and RPSBs (blue open dots). \textbf{Left}: Fitted time since peak of starburst ($t_\mathrm{burst}$) plotted against burst mass fraction ($f_\mathrm{burst}$). \textbf{Right}: Fitted time since peak of starburst ($t_\mathrm{burst}$) plotted against the starburst's quenching timescale as measured by its SFR halving timescale ($\tau_{1/2}$). For both panels, histograms that show the 1D distributions of the x- and y-axes parameters for both PSB types are plotted in the side windows. The PSB regions in both PSB types are very similarly distributed.}
    \label{fig:CPSBvsRPSB_scatter_SFH}
\end{figure*}

In Figure \ref{fig:CPSBvsRPSB_scatter_SFH}, we plot the lookback time to the peak of the starburst ($t_\mathrm{burst}$) against the burst mass fraction ($f_\mathrm{burst}$) and the starburst's halving timescale ($\tau_{1/2}$, time taken for the starburst's SFR to drop to half of the peak value) for both PSB types. The dots mark the posterior median values while the error bars mark the $1\sigma$ regions. 
The strong positive correlations between $t_\mathrm{burst}$ and $f_\mathrm{burst}$ (left panel) and between $t_\mathrm{burst}$ and $\tau_{1/2}$ (right panel) observed for both PSB types can mostly be explained by selection effects, which we discuss in Appendix \ref{apx:SFH}. 
Thus, we do not deduce any conclusions from these correlations, but instead focus on comparing the distributions of CPSBs and RPSBs. 

In the left panel of Figure \ref{fig:CPSBvsRPSB_scatter_SFH}, the PSB regions of CPSBs and RPSBs show similar distributions in $f_\mathrm{burst}$ and $t_\mathrm{burst}$, with the difference between the median values less than or similar to the typical $1\sigma$ estimation uncertainty. The RPSB sample has a slightly higher fraction with high burst age ($t_\mathrm{burst}>2\;$Gyr), but the small number of objects mean this is unlikely to be significant.

In the right panel of Figure \ref{fig:CPSBvsRPSB_scatter_SFH}, most PSBs have a typical $\tau_{1/2}\approx300\;$Myr, in agreement with the quenching timescales estimated by \citet{wild2010} using different methods, but longer than \cite{french2018a} likely due to a difference in the spectral fitting approach. 
The value is also consistent with the timescales of molecular gas depletion in PSBs \citep{rowlands2015,li2019}. 
Comparing the PSB samples, the PSB regions in RPSBs appear to have slightly longer quenching timescale, but the difference is only slightly higher than the typical $1\sigma$ uncertainty of $\tau_{1/2}$ (difference between median $\tau_{1/2}$ is $\sim90\;$Myr, uncertainty $\sim50\;$Myr). 

The starbursts in the PSB regions of RPSBs appear to have comparable ages, strengths and quenching timescales to their CPSB counterparts. To verify this we performed a Monte Carlo Kolmogorov–Smirnov (KS) test on the parameters $t_\mathrm{burst}$, $f_\mathrm{burst}$ and $\tau_{1/2}$. For each parameter, we randomly drew a sample from the parameter's posterior distribution for each galaxy in each PSB type. Then, the two-sample KS test was performed by comparing the randomly drawn samples from CPSBs to those from RPSBs, and the p-value was recorded. This process was repeated $10^5$ times to thoroughly sample the posterior distributions of all galaxy parameters. We found the median p-value of $t_\mathrm{burst}$, $f_\mathrm{burst}$ and $\tau_{1/2}$ were 0.723, 0.550 and 0.127, respectively, indicating that the null hypothesis where the two samples are drawn from the same parent sample could not be rejected for any of the three SFH parameters (adopting $p<0.05$ as rejection of the null hypothesis). Thus, we formally conclude our CPSB and RPSB samples show similar SFHs in their PSB regions.

Comparing the PSB types' dust attenuation, the median fitted $A_V$ of PSB regions in RPSBs is $\sim0.1\;$mag higher than the PSB regions in CPSBs. Although the difference is not significantly larger than the typical $1\sigma$ fitting uncertainty ($\sim0.04\;$mag), the Monte Carlo KS test returns a high statistical significance (median p-value $=0.0061$). Galaxies with higher dust content and dust lanes have been related to recent mergers \citep{goudfrooij1998,shabala2012}, which suggests that RPSBs could either have a higher fraction of post-mergers than CPSBs, or are observed sooner after their mergers, but this would require further analysis of low surface brightness features to confirm.

\subsection{The PSB regions of most RPSBs also increase in stellar metallicity during the starburst} \label{sec:results2}

\begin{figure*}
    \centering
    \includegraphics[width=0.95\textwidth]{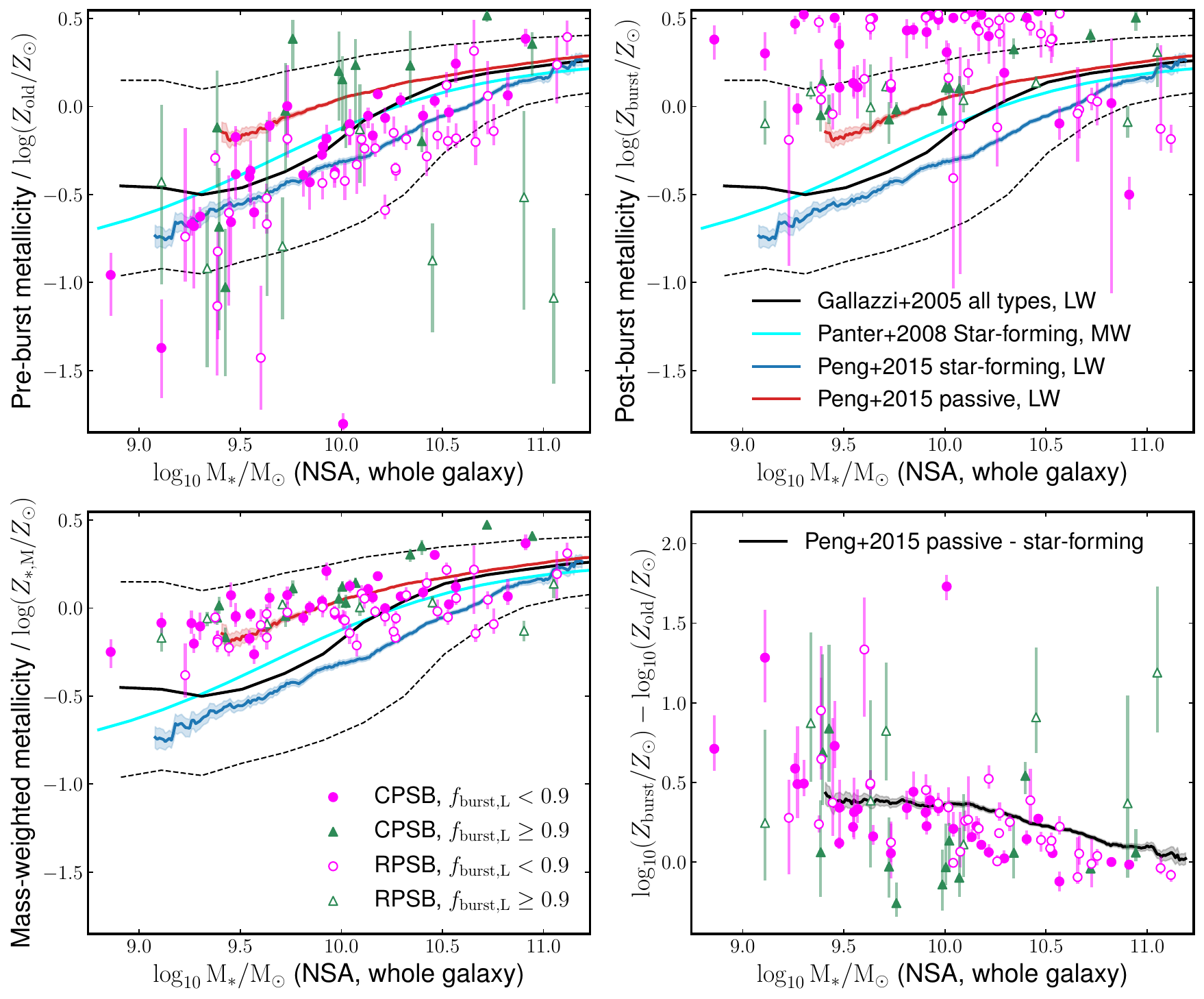}
    \caption{Stellar mass-metallicity relations of the PSB regions in CPSBs (closed symbols) and RPSBs (open symbols), both before \textbf{(upper left)}, during the starburst \textbf{(upper right)} and overall mass-weighted \textbf{(bottom left)}. PSBs with a less dominating burst light fraction (posterior median $f_\mathrm{burst,L}<0.9$) and a higher burst light fraction (posterior median $f_\mathrm{burst,L}\geq 0.9$) are marked with magenta dots and dark green triangles, respectively. All values are plotted against the estimated stellar mass of the whole galaxy from the NSA catalogue. Stellar mass-metallicity relations from the literature are also plotted for comparison, as indicated in the legends. The dashed black lines mark the 16th and 84th percentiles from \protect\cite{gallazzi2005}. The \textbf{bottom right} panel compares the difference between overall mass-weighted and pre-burst metallicity and the difference between passive and star-forming relations from the literature. The PSB regions in RPSBs exhibit very similar patterns as their counterparts in CPSBs, but generally have mass-metallicity relations that are slightly offset to lower metallicity in all three relations, consistent with expected metallicity radial gradients.}
    \label{fig:CPSBvsRPSB_zmet_vs_Mstar}
\end{figure*}

In \cite{paper1}, by placing the pre-burst, post-burst\footnote{Stellar metallicity of stars formed after the burst age $t_\mathrm{burst}$, excludes contribution from stars formed prior and is not a time-averaged measurement.} and overall mass-weighted stellar metallicities of the PSB regions of CPSBs on the mass-metallicity diagram alongside mass-metallicity (MZ) relations of local galaxies from the literature, we showed that the metallicity evolution experienced by the CPSBs during the recent starburst is entirely consistent with the difference between the MZ relations of star-forming and passive galaxies. This result holds for the PSB regions in RPSBs. Figure \ref{fig:CPSBvsRPSB_zmet_vs_Mstar} plots the three aforementioned metallicities against the galaxy's total stellar mass for both CPSBs and RPSBs. Also shown in the first three panels are the MZ relations from three studies of local galaxies: star-forming SDSS galaxies, mass-weighted \citep{panter2008}; SDSS, all types, light-weighted \citep{gallazzi2005}; star-forming and passive SDSS galaxies, light-weighted \citep{peng2015}\footnote{We note that the \cite{peng2015} metallicity estimates are from \cite{gallazzi2005}, but split into star-forming and passive populations.}. 

In the top left panel of Figure \ref{fig:CPSBvsRPSB_zmet_vs_Mstar}, before the recent starburst, the PSB regions of all PSBs broadly followed the known MZ relation where metallicity increases with mass, especially when we consider only the more reliable lower burst light fraction galaxies\footnote{We place less emphasis on the metallicity evolution of PSBs with high burst light fraction because heavy outshining of their older population's light contribution lead to substantial uncertainties in their pre-burst metallicities.} (magenta closed and open dots for CPSB and RPSB respectively). 
This indicates that prior to the starburst, the stellar populations in both types of PSB regions exhibited no atypical recent chemical evolution. 

In the top right panel of Figure \ref{fig:CPSBvsRPSB_zmet_vs_Mstar}, after the recent starburst, significantly more metal-rich stars have formed in most PSB regions in both PSB types (at $>1\sigma$ significance; CPSB: 69\%, RPSB: 63\%), and no longer follow any observable correlation with stellar mass, suggesting that the recent starburst in both PSB types have disrupted the MZ relation. 
Throughout the metallicity evolution, the PSB regions in RPSBs are on average less metal rich than their CPSB counterparts. Compared to CPSBs, new stars formed in the PSB regions of RPSBs are $\approx0.15\;$dex lower pre-burst, and $\approx0.09\;$dex lower post-burst. The differences are comparable to our typical $1\sigma$ estimation uncertainties and have low statistical significance from the Monte Carlo KS test (pre-burst metallicity median $p=0.064$; post-burst metallicity median $p=0.275$). By assuming that the metallicities measured for the CPSBs can be treated as the metallicity at radial distance $r=0\;\mathrm{R}_e$, and the metallicities measured for the RPSBs can be treated as the metallicity at the mean radial distance of all RPSB spaxels in their galaxies $r=1.14\;\mathrm{R}_e$, we estimate that the pre-burst (post-burst) metallicity gradient is $\approx-0.13\;\mathrm{dex/R}_e$ ($\approx-0.08\;\mathrm{dex/R}_e$). These gradients are comparable with the average metallicity gradient measured from early- and late-type local galaxies ($-0.18 \pm 0.02\;\mathrm{R}_e$, \citealt{parikh2021}), thus suggesting that any metallicity differences between the two PSB types are a result of radial gradients.

Galaxies typically have shallower gravitational potential in their outer regions. Thus it might appear surprising that similar to those in CPSBs, the outer PSB regions in most RPSBs formed significantly more metal-rich stars than before their starbursts (Figure \ref{fig:CPSBvsRPSB_zmet_vs_Mstar}). A positive correlation between local stellar mass surface density and metallicity have been reported (stellar: \citealt{neumann2021,zibetti2022}, gas-phase: \citealt{barrera-ballesteros2017,gao2018,baker2023}), thus one could expect a less significant increase in stellar metallicity in the PSB regions of RPSBs compared to CPSBs. We compared our PSB regions' post-burst and overall mass-weighted metallicities as a function of their local stellar mass surface density against the relation reported in \cite{zibetti2022}. Although a sizeable fraction of RPSBs have estimated post-burst metallicities exceeding the distribution of regions in \cite{zibetti2022}, we observe the overall mass-weighted metallicities of both CPSBs and RPSBs are consistent with the relation, in spite of the relation's light-weighted nature. This indicates the significant increase in the metallicity of newly formed stars from pre-burst to post-burst in PSB regions is not strong enough to cause the region to become abnormally metal rich overall for its stellar mass surface density.

Metallicity in a galaxy is regulated by three processes: enrichment through the releasing of metals formed within stars following star formation, dilution through metal-poor gas inflows and the removal of metals through outflows. The first increases metallicity, while the latter two decreases it \citep[e.g.][]{chisholm2018}. Therefore, the significant increase from pre-burst to post-burst metallicity in most PSB regions indicates the starbursts were not driven by a significant inflow of metal-poor gas. Instead, it could be caused by star formation from recycled gas that are more metal-rich, or an increase in star-formation efficiency such as during galaxy mergers \citep{di_matteo2007}, which leads to compact star formation and rapid enrichment.

In the bottom left panel of Figure \ref{fig:CPSBvsRPSB_zmet_vs_Mstar}, we show the overall mass-weighted metallicity plotted against total stellar mass. Here, the PSB regions of CPSBs closely follow the passive light-weighted MZ relation from \cite{peng2015}, while the RPSBs exhibit a similar slope but offset to slightly lower metallicity. As explained in \cite{paper1}, our comparison of mass-weighted to light-weighted MZ relations should not affect our conclusions. 

Lastly, in the bottom right panel, we plot the difference between each galaxy's overall mass-weighted metallicity and the pre-burst metallicity against their total stellar mass, alongside the difference between the passive and star-forming MZ relations from \cite{peng2015}. Once the less reliable $f_\mathrm{burst,L}>0.9$ galaxies are excluded (green triangles), the metallicity difference in the PSB regions in both CPSBs and RPSBs decreases with increasing total stellar mass, and are consistent with the difference reported in \cite{peng2015}. Thus, the results from Figure \ref{fig:CPSBvsRPSB_zmet_vs_Mstar} show that, regardless of whether the PSB regions are distributed in the galactic centre or in a ring formation at larger radial distances, the starburst followed by rapid quenching in a PSB phase can result in a significant metal enhancement in these regions, a metallicity change that is consistent with the large gap found between the star-forming and passive MZ relations reported in the literature.

\subsection{The SFHs of the outer PSB regions in RPSBs are closely tied to their central regions} \label{sec:RPSBvsCentral_results}

\begin{table*}
    \centering
    \caption{Posterior estimated properties of the central regions of 41 RPSBs from the spectral fitting of stacked MaNGA spaxels. The central regions of 4 galaxies marked by dashes were poorly fit and are not considered in the analysis in Section \ref{sec:RPSBvsCentral_results}. Columns are (1) MaNGA Plate-IFU,
    (2) stellar mass within the stacked central spaxels,
    (3) ISM dust attenuation at 5500\AA\ ($V$ band),
    (4) the $\log_{10}$ SFR within the stacked central spaxels averaged over the last 100\,Myr,
    (5) the corresponding $\log_{10}$ sSFR averaged over the last 100\,Myr,
    (6) the mass-weighted age,
    (7) SFR halving timescale since the peak SFR,
    (8) time since the peak SFR,
    (9) stellar metallicity,
    and (10) reduced chi-squared value of the maximum likelihood posterior sample spectrum. 
    The full table is available as supplementary online material.}
    \begin{tabular}{llllllllll}
        \hline
        Plate-IFU & $\log_{10}$ & $A_V$/mag & $\log_{10}\mathrm{SFR_{100Myr}}$ / & $\log_{10}\mathrm{sSFR_{100Myr}}$ / & $t_M$/Gyr & $\tau_{1/2}$/Myr & $t_\mathrm{peak}$/Gyr & $Z_*/Z_\odot$ & $\chi^2_\nu$ \\
        (1) & $M_{*,\mathrm{central}}/M_\odot$ (2) & (3) & $\mathrm{M_\odot\; yr^{-1}}$ (4) & $\mathrm{yr^{-1}}$ (5) & (6) & (7) & (8) & (9) & (10)\\
        \hline
        7990-12705 & $10.39^{+0.03}_{-0.02}$   & $1.69^{+0.02}_{-0.02}$    & $-1.01^{+0.13}_{-0.11}$   & $-11.39^{+0.11}_{-0.12}$  & $3.34^{+0.57}_{-0.42}$    & $301^{+34}_{-35}$         & $1.76^{+0.10}_{-0.11}$    & $0.72^{+0.03}_{-0.03}$    & 0.864  \\
        8080-3704  & $9.67^{+0.01}_{-0.01}$    & $1.16^{+0.02}_{-0.02}$    & $-0.43^{+0.02}_{-0.02}$   & $-10.09^{+0.02}_{-0.02}$  & $2.82^{+0.20}_{-0.13}$    & $410^{+27}_{-26}$         & $1.30^{+0.05}_{-0.06}$    & $1.20^{+0.05}_{-0.06}$    & 0.973  \\
        8083-12703 & $9.74^{+0.01}_{-0.01}$    & $0.77^{+0.02}_{-0.02}$    & $0.25^{+0.02}_{-0.02}$    & $-9.49^{+0.01}_{-0.02}$   & $1.41^{+0.08}_{-0.06}$    & $1839^{+222}_{-152}$      & $2.00^{+0.13}_{-0.12}$    & $1.91^{+0.06}_{-0.07}$    & 0.877  \\
        8146-1901  & $10.01^{+0.01}_{-0.01}$   & $1.09^{+0.02}_{-0.02}$    & $0.68^{+0.01}_{-0.01}$    & $-9.33^{+0.01}_{-0.01}$   & $1.13^{+0.06}_{-0.04}$    & $1981^{+219}_{-191}$      & $1.65^{+0.09}_{-0.09}$    & $2.41^{+0.03}_{-0.04}$    & 0.550  \\
        8152-3703  & $10.13^{+0.01}_{-0.01}$   & $0.75^{+0.02}_{-0.02}$    & $-0.59^{+0.03}_{-0.04}$   & $-10.72^{+0.04}_{-0.03}$  & $6.78^{+0.15}_{-0.42}$    & $611^{+121}_{-70}$        & $3.27^{+0.40}_{-0.32}$    & $0.55^{+0.03}_{-0.03}$    & 0.990  \\
        \multicolumn{9}{c}{$\ldots$} \\
        \hline
    \end{tabular}
    \label{tab:RPSB_central_results}
\end{table*}

In Table \ref{tab:RPSB_central_results}, we report the properties derived from our fits to the central regions of our 41 RPSBs. This includes the regions' stellar mass, the $V$-band dust extinction strength ($A_V$), the SFR and sSFR calculated over the past 100 Myr period, the mass-weighted stellar age, metallicity, and reduced chi-squared of the maximum likelihood posterior sample spectrum. 

The lower panel of Figure \ref{fig:RPSB_fit_example} shows an example fit of the central region of a RPSB, the same galaxy to match the PSB region in the upper panel. In both regions, the posterior median model spectrum (orange) traces the stacked observed spectrum (black) well outside of the masked regions (light grey bands). The reduced chi-squared value noted for both regions are close to unity, indicating a good fit. Comparing the fitted GP noise component of the central region to its outer PSB region, we observe that the majority of the corrections introduced by the GP noise are present in both, suggesting a common origin from issues related to the galaxy's observation or calibration. However, we also observe the central region's fit has a stronger positive contribution at the blue, UV end compared to the PSB region (observed wavelength
$\approx 3700-4300\ \text{\normalfont\AA}$). This blue GP continuum is found in at least half of the central regions fitted, and is possibly caused by the GP noise correcting for systematics in the stellar population synthesis templates.

The blue GP continuum could be caused by the incomplete modelling of the spectral contribution from hot evolved stars, where current stellar population synthesis models lack a blue component to accurately match observed spectra of low-metallicity horizontal branch stars \citep[see][Section 6]{ocvirk2010,cid-fernandes2011}. \cite{ocvirk2010} have shown that this artifact is widespread and can contribute up to 12\% of the integrated light in optical wavelengths. Additionally, the artifact could be a result of the \cite{bruzual2003} models not including the effects of binary stars. \cite{eldridge2017} and \cite{stanway2018} have shown that when the effects of binary stars are considered in stellar population synthesis models, the model spectra of $>1\;$Gyr stellar populations can be considerably bluer. The blue GP continuum could therefore be compensating for the redder model spectra from the \cite{bruzual2003} models that have incorrect horizontal branch models or lack binary stars. Considering we do not observe significant differences in the estimated SFHs reported here and those from the ``continuity'' SFH model without GP noise, we do not expect the uncertain modelling of the blue continuum to affect our results and conclusions.

\subsubsection{Outer PSB regions exist in globally quenching galaxies} \label{sec:centralSFH}
\begin{figure*}
    \centering
    \includegraphics[width=\textwidth]{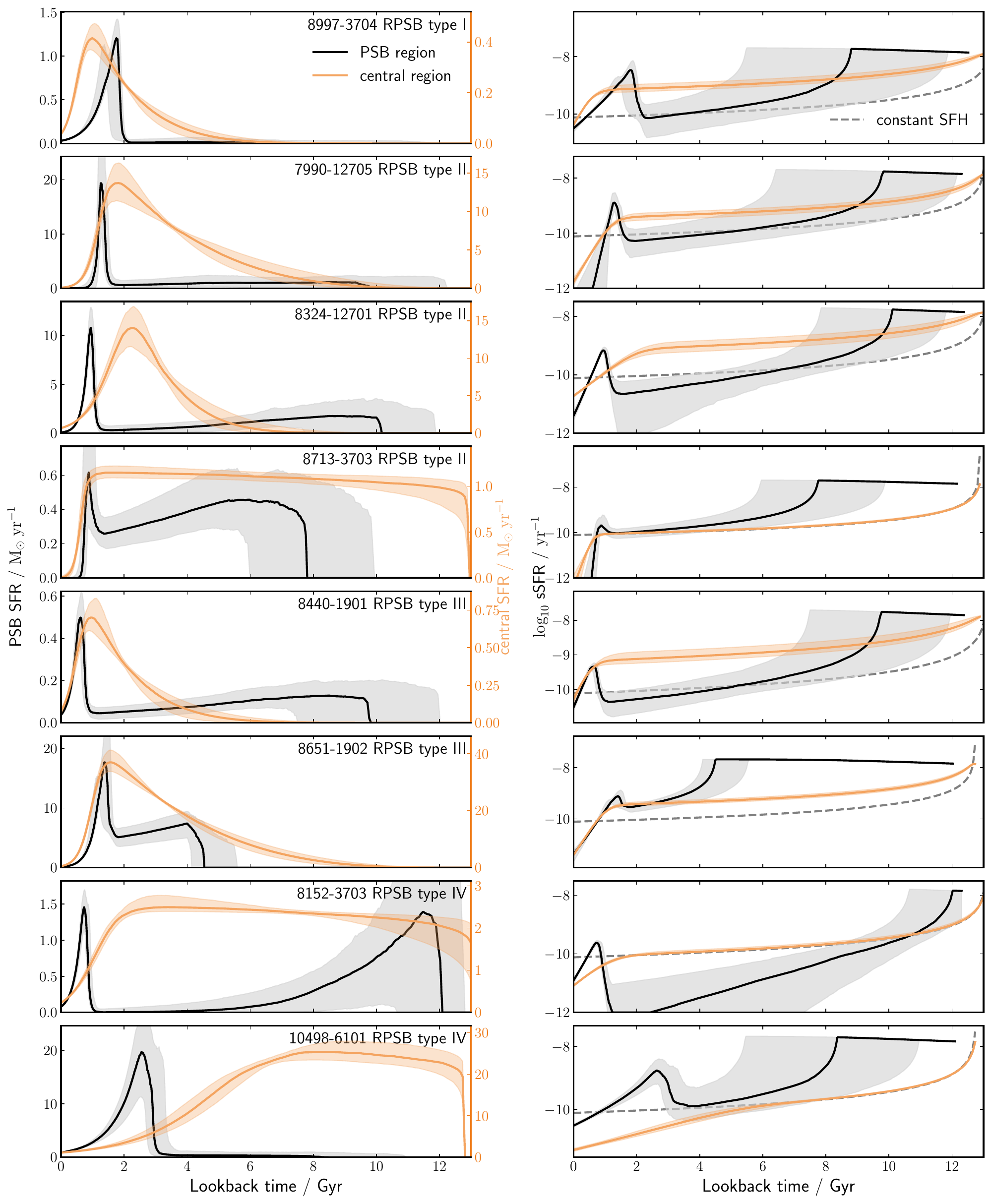}
    \caption{The fitted SFHs (left) and sSFR evolution of the PSB (black) and central (orange) regions of 8 example RPSBs. Note that in the left column of panels, the PSB and central regions are plotted on different y-axes. The shaded regions mark the $1\sigma$ uncertainty on all SFHs and sSFR traces. In the right panels, the dashed lines mark the sSFR evolution of a constant SFH, given by $1/t_H$ where $t_H$ is the age of the Universe (Hubble time) in years.}
    \label{fig:RPSBvsCentral_sfh}
\end{figure*}

In the left column of Figure \ref{fig:RPSBvsCentral_sfh}, we present the fitted SFHs from the central and PSB regions of 8 example RPSBs. In addition, in the right column we compare the two regions' SFHs via their sSFR evolution. The figures for the other 29 RPSBs with successful fits for both their central and PSB regions are provided in Appendix \ref{apx:RPSB_sfhs}. From both the SFR and sSFR of the RPSBs, we found no RPSB with a central region that is rising in SFR, and only one RPSB with a central region that has been plausibly maintaining a constant SFR (9891-6102). Nearly all central regions are quenching within the last 2 Gyr: the mean estimated sSFR at a lookback time of 2 Gyr is $-9.23\pm0.13\;\mathrm{yr}^{-1}$; the mean estimated current sSFR is $-10.24\pm0.09\;\mathrm{yr}^{-1}$. This $\sim1.0\;$dex difference is considerably larger than the decline in sSFR of main sequence star-forming galaxies over the same period ($\approx0.2\;$dex, \citealt{popesso2023}). 

The previous comparison to main sequence galaxies could be biased by comparing our central sSFR decline to the sSFR decline measured from entire galaxies. Therefore, to provide a more direct comparison, we construct a control sample of central regions drawn from all non-PSB galaxies in the full MaNGA DR17 sample. For each of the 37 RPSB with successful fits to their central regions, we randomly draw 10 non-repeating mass- and redshift-matched controls from the full sample, and obtain a emission-line subtracted 1D spectrum stacked from its central region matched to the corresponding RPSB's central region in $R/R_e$. We found that at fixed $\mathrm{D}_n(4000)$ break strength, the central regions of RPSBs are typically $\sim0.9\text{\normalfont\AA}$ higher in the $\mathrm{H}\delta_\mathrm{A}$ index than the central regions of controls, thus indicating they have experienced more recent and/or rapid quenching than the wider galaxy population \citep{kauffmann2003a,goncalves2012}.

Despite their quenching trend, these central regions are not selected as PSB regions by our selection methods. This is due to either residual star formation, lack of a short and sharp recent starburst or H$\alpha$ emission from another ionising source such as shocks or AGN, or a combination of all three. As seen from Figure \ref{fig:RPSBvsCentral_sfh}, beyond the common recent SFR decline, the central regions of RPSBs show a large diversity in past evolution. The detailed history of RPSBs will be further discussed in Section \ref{sec:RPSB_origin}.

\subsubsection{Weaker central sSFR is related to quicker outer region quenching}
\begin{figure*}
    \centering
    \includegraphics[width=\textwidth]{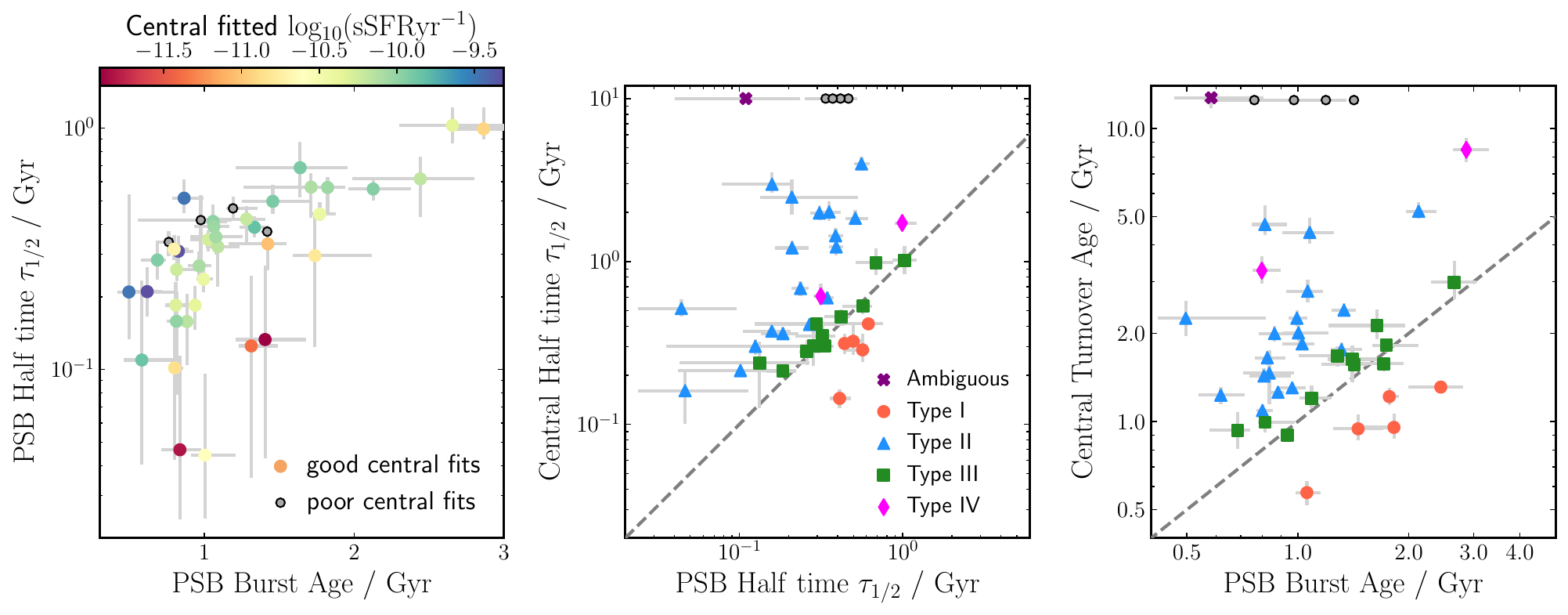}
    \caption{Relations between the recent SFH of central and outer PSB regions in RPSBs. \textbf{Left:} The fitted time since peak of starburst ($t_\mathrm{burst}$) plotted against the starburst's quenching timescale ($\tau_{1/2}$) of the PSB regions, coloured according to the posterior median sSFR value of the same galaxy's central region. All datapoints' location are identical to the RPSB datapoints in the right panel of Figure \ref{fig:CPSBvsRPSB_scatter_SFH}. RPSBs with central regions that failed during our fitting process are shown as smaller gray dots with black rims. \textbf{Middle:} Comparing the quenching timescale of the central and PSB regions, with the different colours and symbols indicating the RPSB types of the galaxies. The dashed gray line marks the $1:1$ relation. \textbf{Right:} Similar style as the central panel, but instead compares the burst age of the PSB region to the turnover age (time since maximum SFR) of the centre. In the middle and the right panels, the y-axis location of the RPSBs with poor central fits are placed arbitrarily, and do not reflect their fitted values.}
    \label{fig:RPSBvsCentral_burstage_vs_halftime}
\end{figure*}

We next investigate the relationships between the SFHs of the outer PSB and the central regions of RPSBs by studying the correlations between properties derived from both regions' SFHs. In the left panel of Figure \ref{fig:RPSBvsCentral_burstage_vs_halftime}, we show the PSB regions' estimated $t_\mathrm{burst}$ against $\tau_{1/2}$ coloured by the posterior median sSFR value of the same galaxy's central region (calculated over the most recent 100 Myr). A weak trend between $\tau_{1/2}$ and central sSFR can be observed. At a fixed burst age, RPSBs with a central region at lower sSFR tend to have experienced more rapid quenching in the recent starburst in their PSB regions.

In the middle and right panels of Figure \ref{fig:RPSBvsCentral_burstage_vs_halftime}, we compare the quenching timescale and time since the most recent peak in SFR in the fitted SFHs of the central and PSB regions. The different symbols and colours denote the RPSB types of the galaxies, which is discussed in detail in Section \ref{sec:RPSB_origin}. In both panels, a large fraction of the RPSBs appear to follow a positive correlation, where galaxies with more rapid and recently quenching central regions also possess outer regions that underwent more rapid and recent quenching. The correlation in the middle panel is stronger, potentially suggesting a stronger link in quenching timescales across the regions.

Therefore, Figure \ref{fig:RPSBvsCentral_burstage_vs_halftime} indicates that for RPSBs with central regions that have quenched earlier or quicker to lead to a low current sSFR, their outer PSB region will also likely have experienced rapid quenching. 
On the other hand, for RPSBs that have a continuously star-forming core (higher sSFR and longer quenching timescales), their outer PSB region's recent starburst will have experienced slower quenching.

However, we note that the above comparisons are made assuming the central and outer PSB regions can be observed independently. This is not true for galaxies that are not face-on, where the spectrum from a part of the outer region, likely PSB, is superimposed onto the spectrum from the brighter centre. Thus, PSB-like signatures could be introduced to the central stacked spectrum, artificially boosting correlations between the two regions' SFHs. To test the degree of this projection effect, we estimated the RPSBs' inclination from their axial ratios ($b/a$), and compared the integrated flux density of the PSB regions to their central regions. The galaxies with the lowest axial ratios and highest PSB to central flux ratios did not show more closely matching fitted stellar properties between their central and PSB regions compared to other RPSBs, and do not preferably lie on the $1:1$ line in the middle and right panels of Figure \ref{fig:RPSBvsCentral_burstage_vs_halftime}. Therefore, we do not consider projection effects to have produced the co-evolution signatures seen between the central and outer PSB regions in our RPSB sample.

\section{Discussion} \label{sec:CPSBvsRPSB_discussion}
Our results in Sections \ref{sec:results1} and \ref{sec:results2} show that despite the differences in location in their galaxies, the PSB regions in CPSBs and RPSBs exhibit very similar properties in SFH and metallicity evolution. Focusing on RPSBs only, our results in Section \ref{sec:RPSBvsCentral_results} show that all their central regions are experiencing a decline in SFR, and that the outer PSB regions' recent SFH are closely linked with the centre's SFH. In this section, we discuss the implications of our results on the quenching mechanisms involved in rapid star formation quenching.

\subsection{Common local driving mechanisms in rapidly quenched regions regardless of quenching trigger}

Our results show that PSB regions have remarkably similar SFH and chemical properties, regardless of their location in their galaxy. Thus, even if different driving mechanisms lead to differing spatial distributions of the starburst and subsequent rapid quenching, a similar set of secular processes at the local scale seem to result in a similar imprint on the star formation and chemical history. This result is supported by findings regarding the gas properties in PSBs. \cite{otter2022} analysed the molecular and ionized gas properties of a small sample of 13 MaNGA PSBs, finding that CPSBs and RPSBs have similar gas fractions for their stellar mass. Regardless of the location of the PSB region, molecular gas mass surface density is elevated compared to SFR surface density, above the expected Kennicutt-Schmidt relation of star-forming galaxies. Thus, the PSB regions distributed in both the centre and outskirts of galaxies are currently inefficient in star formation. 

\cite{smercina2018} and \cite{otter2022} proposed turbulence in the ISM that would otherwise support higher SFRs as the main cause of the PSBs' low current star-formation efficiency, maintaining the galaxies' quiescence after being rapidly quenched by prior events. ISM turbulence can be caused by nuclear events such as AGN feedback \citep{piotrowska2022} or disturbances to the gravitational potential following a merger \citep{scoville2017}, and events at larger radial distance such as a rotating bar \citep{salim2020b} or stellar feedback. Therefore, it is possible that turbulence is the common local-scale process regulating the starburst and rapid quenching in PSB regions, regardless of location within the galaxy or triggering mechanism. We turn to the question of what the prior events could be driving this turbulence in the next subsections. 

\subsection{The diverse evolutionary pathway of RPSBs} \label{sec:RPSB_origin}
While inner PSB regions are generally thought to be caused by mergers leading to torques driving gas to the centre, which drives the central starburst \citep[e.g.][]{hopkins2006,chen2019}, it is not clear whether strong AGN feedback is also required to completely quench star formation in the centres \citep{zheng2020}. For the first time we can look at the causes of outer PSB regions, which will help to constrain the mechanisms responsible for rapid quenching in general.

Examining the SFH of the central and PSB regions of our RPSB sample (8 examples are shown in Figure \ref{fig:RPSBvsCentral_sfh}, others in Appendix \ref{apx:RPSB_sfhs}), we see that prior to the recent decline in central SFR in the most recent $\sim2\;$Gyr, the RPSBs exhibit diverse SFHs in both their central and PSB regions. \cite{chen2019} already suggested that RPSBs have diverse origins. Using the spaxels' location in the $\mathrm{H\delta_A}$ - $\mathrm{W(H\alpha})$ plane, the authors divided their sample of RPSBs into two types. Type I are galaxies where the outer PSB regions began quenching earlier than the central regions, both with similar quenching timescales, or are galaxies that have less complete central quenching. Type II are galaxies that began quenching globally at around the same time, but the outer PSB region has quenched significantly quicker. We see clear examples for both types in our reconstructed SFHs. 

\begin{figure*}
    \centering
    \includegraphics[width=0.83\textwidth]{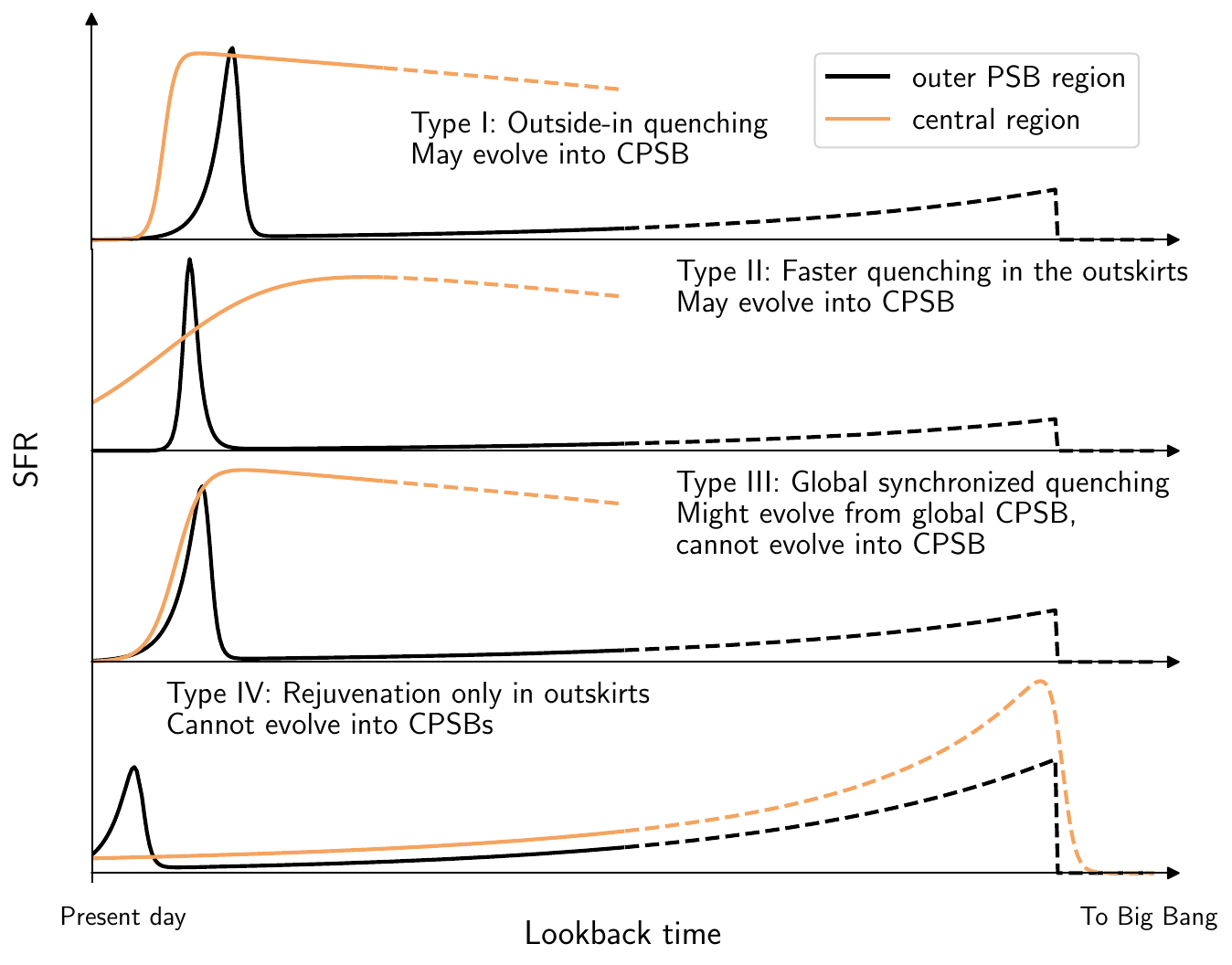}
    \caption{Illustrative examples of typical SFHs of the four RPSB evolutionary pathways. \textbf{Type I} describes an outside-in sequence of quenching, with comparable quenching timescales. When the outer region is selected as a PSB, the central region might remain star forming or only recently began quenching. \textbf{Type II} describes quenching that began in unison at all radial distances or earlier in the centre, but the outer region quenched more quickly. Thus, considerable residual star formation remains in the centre at the time of observation. \textbf{Type III} describes global synchronized quenching, where all radial distances quenched at comparable times with comparable quenching timescales. The reason the centre is not classified as PSB in some selection methods is due to a period of continuous SFR before quenching, unlike the outskirts' starburst. \textbf{Type IV} describes rejuvenation limited to the outer regions, where the centre has long since quenched or began quenching before the outskirts' starburst. The dashed and missing periods of SFH indicates lookback times that are less important to the RPSB's classification.}
    \label{fig:RPSB_type_sketch}
\end{figure*}

Moreover, we identify two additional RPSB types: global synchronous quenching, and rejuvenation in the outskirts (types III and IV, labelled in Figure \ref{fig:RPSBvsCentral_sfh}). Our proposed evolutionary pathways are summarized in Figure \ref{fig:RPSB_type_sketch}, and we provide the number of galaxies in each type we classified through visual inspection of the SFHs in Table \ref{tab:RPSB_numbers}. We do not classify 9891-6102 due to its ambiguous central SFH suggesting it might not be actively quenching. Below, we describe each type in detail.

\begin{table}
    \centering
    \caption{The number of galaxies in our RPSB sample classified into the four RPSB types through visual inspection of their fitted RPSB and central region SFHs. One galaxy, 9891-6102, is excluded due to its ambiguous central SFH suggesting it might not be actively quenching.}
    \begin{tabular}{llll}
         \hline
         Type I & Type II & Type III & Type IV \\
         5/37 & 18/37 & 11/37 & 2/37 \\
         \hline
    \end{tabular}
    \label{tab:RPSB_numbers}
\end{table}

\textbf{Type I}: We found 5/37 RPSBs with SFHs that show outside-in quenching. One example, 8997-3704, is shown in Figure \ref{fig:RPSBvsCentral_sfh}. As seen in the left panel, the outer PSB region peaked in star formation and began quenching at least 250 Myr before the galaxy's central region, with a comparable declining rate in sSFR between the two (right panel). This galaxy was also used as an example for Type I by \cite{chen2019}. Type I RPSBs are found below the $1:1$ lines in the middle and right panels in Figure \ref{fig:RPSBvsCentral_burstage_vs_halftime}, due to their PSB regions' earlier peak in SFR and slightly faster quenching. Their central regions are not classified as PSB due to residual star formation causing higher $\mathrm{W(H\alpha)}$.

\textbf{Type II}: We found a large fraction (18/37) of RPSBs with SFHs that show more rapid quenching in their outer PSB regions compared to their central regions. Examples shown in Figure \ref{fig:RPSBvsCentral_sfh} are 7990-12705, 8324-12701 and 8713-3703. The quenching began either at the same time in both regions (e.g. 8713-3703) or earlier in the centre (e.g. 8324-12703). These galaxies are best identified through the sSFR plots in the right column in Figure \ref{fig:RPSBvsCentral_sfh}, where their PSB regions' decline in sSFR are significantly steeper than their central counterparts. They generally lie above the $1:1$ lines in the middle and right panels of Figure \ref{fig:RPSBvsCentral_burstage_vs_halftime}, indicating slower and earlier quenching in their centres. By selection, these galaxies have rapidly quenched PSB regions, with a quenching timescale of $\tau_{1/2}\approx250\;$Myr.

\textbf{Type III}: We found 11/37 RPSBs with SFHs that show the central and outer regions underwent a synchronized, global quenching event with similar quenching timescales. Two examples of this RPSB type are shown in Figure \ref{fig:RPSBvsCentral_sfh} (8440-1901 and 8651-1902). These galaxies are identified by their central and outer PSB regions beginning quenching at the same time in the left panels (within $1\sigma$ uncertainties) and both regions declining in sSFR at similar rates in the right panels (within $1\sigma$ uncertainties). Thus, they are found close to the $1:1$ lines in the middle and right panels of Figure \ref{fig:RPSBvsCentral_burstage_vs_halftime}.

\textbf{Type IV}: We found 2/37 RPSB, 8152-3703 and 10498-6101, with a clear rejuvenation in the outer PSB regions as shown in Figure \ref{fig:RPSBvsCentral_sfh} (although 10498-6101 could also be identified as inside-out formation). These galaxies' central region have begun their quenching significantly earlier than the outer PSB regions, gradually dropping below the sSFR trace of a constant SFH (dashed line, right panels). During the central regions' quenching, the galaxies' PSB regions remained very low in SFR until the more recent starburst, which drives the PSB regions' sSFR back above the sSFR trace of a constant SFH. This starburst is not accompanied by any increase in central SFR indicating that the rejuvenation is limited to the outer regions.

Given that our results show most of the central regions of the RPSBs have recently quenched, it is interesting to consider why the central regions were not selected as PSB. Regions that rapidly quench following relatively constant star formation (Type III) can be identified as PSB in our selection scheme, but for a much shorter time than galaxies that underwent a starburst prior to quenching, due to a weaker excess in A- and F-type stars. Thus, these galaxies might first be identified as CPSBs with global PSB regions, and are then selected as RPSBs as the inner $\mathrm{H\delta}$ absorption weakens below our threshold. We note that different PSB classification methods might identify these central regions as PSB. For example, using the PCA method \citep{wild2007}, \cite{otter2022} classified 8080-3704 as a CPSB (see our fitted SFHs of this galaxy in Appendix \ref{apx:RPSB_sfhs}). Additionally, a central AGN could boost the $\mathrm{W(H\alpha)}$ of the central region leading to misclassification as a non-PSB using our selection method, but not the PCA method of \cite{wild2007}. 

A corollary to this is that, while we have provided fractions of RPSBs that experienced the four evolutionary pathways in our sample, the fractions should not be directly interpreted as the true underlying fractions of quenching pathways in rapidly quenched galaxies as they are dependent on selection method\footnote{Our PSB samples are selected through the combination of weak H$\alpha$ emission and strong $\mathrm{H\delta}$ absorption, which although have a long visibility window in selecting PSBs after they begin quenching, have a considerably shorter visibility window for galaxies or regions that have not experienced a starburst prior to rapidly quenching. Our selection will miss rapidly quenched objects without a preceding starburst or those with residual star formation.}.
A rigorous investigation into the relative proportions of rapid quenching pathways should select galaxies which have undergone recent rapid quenching directly from their recovered SFHs. Unfortunately, this would require significant computational time to fit all 10,000 MaNGA galaxies in search of a complete sample of regions of rapidly quenching SFHs. 

To summarize, from comparing the fitted SFHs of the central and outer PSB regions of our RPSB sample, we identify four distinct evolutionary pathways: outside-in quenching (5/37), more rapid quenching in the outskirts (18/37), global synchronous quenching (11/37) and outer rejuvenation (2/37). 

\subsection{Will RPSBs become CPSBs, or vice-versa?}
Previous studies have investigated the evolutionary connection between CPSBs and RPSBs \citep{chen2019,cheng2024}. Here, we discuss whether each of our hypothesized RPSB types can evolve into or has evolved from a CPSB:
\begin{itemize}
    \item Type I (5/37): The central regions are typically not selected as PSB due to significant $\mathrm{W(H\alpha)}$. If the centre continues on their current rapid quenching and remain quenched, it is likely they evolve into CPSBs.
    \item Type II (18/37): The central regions are not selected as PSB due to residual star formation in the centre. If the centre quenches completely and rapidly enough to cause the H$\delta$ to strengthen, these may also evolve into CPSBs.
    \item Type III (11/37): These central regions are not classified as PSB due to their extended period of star formation before quenching. These RPSBs could have been CPSBs closer to when quenching began, but they cannot evolve into CPSBs.
    \item Type IV (2/37): These central regions have long since quenched. They cannot evolve into a CPSB without a new episode of central star formation.
\end{itemize}

Our results suggest that the outside-in migration of PSB regions could occur in about half of our RPSB sample (23/37, I+II), with only 5/37 likely candidates. Combined with the Type III that could have been CPSBs in the past, we are unable to exclude an evolutionary link between RPSBs and CPSBs in the majority of cases, either through RPSBs evolving into CPSBs or vice-versa. Our low fraction of likely RPSB to CPSB transformation is consistent with \cite{chen2019}, where the authors concluded that most RPSBs cannot evolve into CPSBs due to their vastly different radial profiles in mass-weighted age. Our results are not entirely in agreement with \cite{cheng2024}. From measuring the radial gradients of spectral indices $\mathrm{D_n 4000}$, $\mathrm{W(H\delta_A)}$ and $\mathrm{W(H\alpha)}$, the authors concluded that their sample of MaNGA CPSBs and RPSBs both generally have older stellar ages and lower recent star formation in the outskirts, thus are both experiencing outside-in quenching but are observed at different stages. The finer detail brought to light through detailed SFH analysis again highlights the importance of this approach over averaged quantities such as a mean stellar age or spectral indices for understanding the full complexity of quenching pathways in galaxies.  

\subsection{The mechanisms that cause rapid quenching}
Here, we return to the question posed in Section \ref{sec:intro}: Are the different spatial distributions of PSB regions in PSB galaxies caused by different quenching mechanisms?

Ram-pressure stripping in dense environments is a commonly attributed cause for rapid quenching in outer galaxy regions. It is shown to preferentially halt star formation in the outskirts of in-falling cluster galaxies \citep{poggianti2019,werle2022}, and can promote star formation in the centre \citep{vulcani2020,roberts2022} by driving increasing gas inflow into the centre \citep{akerman2023}. This differential effect on the inner and outer regions of the satellites could explain the observation of galaxies with rapidly quenched outskirts but a more star-forming core, but it is unclear whether it would cause outside-in quenching (RPSB type I) or speed up the quenching of the outskirts (RPSB type II). 
However, through cross-matching our sample with the GEMA-VAC cluster catalogue \citep[][version DR17]{GEMA-VAC}, we have verified that there is no clear correlation between local galaxy density and RPSB type.
Since our sample are predominantly field galaxies, we do not believe ram-pressure stripping to be a major contributor in their quenching.

Galaxy mergers and interactions can provide catastrophically disruptive events that temporarily drive up gas density and thus star formation, followed by rapid quenching once the gas is exhausted or expelled. Mergers have often been suggested to play an important role in causing local CPSBs in the literature. Local CPSBs have been found to have higher fractions with signs of recent merger than controls \citep[e.g.][]{pawlik2018,wilkinson2022}, while high fractions of post-mergers are found to be CPSBs \citep{ellison2022,li2023,ellison2024}. Based on their significant increase in central stellar metallicity during the starburst, we concluded in \cite{paper1} that most CPSBs could be triggered by gas-rich mergers. Due to the similarity in SFH and metallicity increase between CPSBs and RPSBs that underwent global synchronous quenching (RPSB Type III), galaxy mergers are an obvious candidate cause. 

Could mergers be responsible for all types of RPSBs? Slight differences in the location of gas deposition or density increase might plausibly result in differences in PSB regions' spatial distribution. Aside from the ``classic'' blue to red transition that globally-quenched PSBs have experienced, based on mock PSBs in the EAGLE cosmological simulation, \cite{pawlik2019} also discovered a blue to blue cycle as an alternative PSB evolutionary pathway. Following a merger, the authors showed that some simulated PSBs maintained a reduced but detectable SFR after the starburst. If the residual star formation is predominantly located in the centre, while the outskirts enter quiescence, this could lead to RPSBs with more rapid quenching in their outer regions (RPSB type II). The difference in the regions' quenching timescale might be caused by the centre's greater gravitational potential, which more readily accretes the remaining available gas to form stars, or the impact of gravitationally induced torques that efficiently drive gas to the centre. 

From hydrodynamical simulations of binary mergers, \cite{petersson2023} showed that the starburst triggered by a merger's initial flyby could be globally distributed, while the starburst triggered during the later nuclear coalescence is limited to the centre. A small number of PSBs in our sample show outside-in quenching characteristics (RPSB type I), with central SFHs that peak $\sim500\;$Myr after the outer regions. These are consistent with this picture: the initial flyby mainly impacts the outer disc, followed by the later coalescence driving gas inwards to trigger the central starburst and subsequent quenching as gas is exhausted or expelled through feedback.

Finally, rejuvenation only in the outskirts (RPSB type IV) could be caused by accretion of gas by a quenched or near-quenched galaxy that is primarily deposited in the outskirts. This gas could come directly from the circum-galactic medium (CGM) or through a gas-rich minor merger. The newly accreted gas promotes a period of increased star formation in the outskirts, before quenching once the gas is exhausted \citep{pawlik2019}. We only found 2/41 RPSBs with a high confidence of this evolutionary pathway, suggesting that this type of event is relatively rare at low redshift.

It is important to also discuss the impact of centralised feedback processes on the spatial distribution of quenched regions. If gas is driven inwards to the central super-massive black hole, AGN feedback could disrupt star formation via heating or outflows \citep{feruglio2010,cicone2014}. This is generally expected to impact the central regions of the galaxy more than the outer regions \citep[e.g.][]{zheng2020}, and could occur 10's-100's of Myr after the initial disruptive event \citep{wild2010, hopkins2012}. We found the greatest fraction of possible AGN hosts in PSBs with more rapid outer quenching (RPSB type II, 10/18), which might indicate AGNs playing a role in the gradual suppression of their central star formation through weaker ``maintenance" mode feedback \citep{croton2006}. Although we found low fractions of AGN hosts in Type I or III RPSBs (I: 0/5, III: 2/11), due to the high variability of AGN, we cannot rule out AGN feedback as a contributing factor in their rapid decline in central star formation.

The correlation between the central sSFR and outer quenching timescales in RPSBs (Figure \ref{fig:RPSBvsCentral_burstage_vs_halftime}) indicates considerable co-evolution between their central and outer regions. The simplest explanation is that a common gas supply, provided by any of the mechanisms discussed above, is fuelling both central star formation and the outer starburst. If less gas is available to sustain a long duration outer starburst, then less gas is available for the inner region too, thus resulting in RPSBs with more rapidly quenched PSB regions to also have centres with low sSFR. Alternatively, if central AGN feedback is driving the quenching at all radial distances, galaxies with more rapid outer quenching and lower central sSFR could be experiencing more intense AGN feedback.

We conclude that mergers and/or interactions could cause all types of CPSB and RPSB galaxies through different responses of the gas in the galaxy to the disrupted gravitational potential, and the different properties (e.g. mass ratios, velocities, orientations) of the mergers. Detailed hydrodynamical simulations of various quenching mechanisms to produce predictions for resolved SFH would greatly increase our ability to definitively assign physical mechanisms to the diverse observed evolutionary pathways.

\section{Summary and conclusions}
We selected and stacked the post-starburst regions of 50 PSBs with central PSB regions (CPSB) and 41 PSBs with outer PSB regions (RPSB) from the MaNGA IFU survey.
From Bayesian full spectral fitting of the resulting high $\mathrm{SNR}>100$ stacked spectra, we obtained full SFHs of the PSB regions and stellar metallicity both before and after the recent starburst. 
Additionally, we fitted the stacked spectra from the central regions of the 41 RPSBs to obtain the full SFHs of the galaxy cores.

From the final sample of 45 well-fitted CPSBs, 41 RPSBs and 37 central regions, our results lead to five main conclusions:
\begin{enumerate}
    \item The PSB regions of galaxies show very similar distributions of star formation and chemical enrichment properties, regardless as to where they are located within the galaxy (Figures \ref{fig:CPSBvsRPSB_scatter_SFH} and \ref{fig:CPSBvsRPSB_zmet_vs_Mstar}). These similarities suggest that, even if different driving mechanisms cause the starburst and subsequent rapid quenching in inner and outer PSB regions, the same local scale processes regulate gas supply, cooling, feedback, metal supply and quenching, resulting in similar stellar population properties. This is consistent with the idea that excess turbulence in the ISM causes the quenching.

    \item In galaxies with outer post-starburst regions, we find the central regions are also in the process of shutting down their star formation in 36/37 cases (Figure \ref{fig:RPSBvsCentral_sfh}). Their central sSFR has declined by $\sim1.0\;$dex on average in the most recent 2 Gyr, significantly steeper than the decline experienced by main sequence star-forming galaxies over the same period ($\approx0.2\;$dex). This suggests a global catastrophic event as a causation mechanism for all PSBs, regardless of where the PSB region is detected in the galaxy.

    \item We find that the more rapidly the outer region quenches, the lower the sSFR of the inner region, at a fixed burst age (Figure \ref{fig:RPSBvsCentral_burstage_vs_halftime}). This indicates considerable co-evolution between the central and outer regions, which we hypothesise could be caused by both regions being fuelled by a common gas supply and/or impacted by central AGN feedback.

    \item The quenching of PSB galaxies has varied radial patterns, including progressing inside-out, outside-in and synchronous (Figure \ref{fig:RPSBvsCentral_sfh}). Quenching in the inner region can occur more slowly or just as rapidly as the outer region. This diversity in quenching pathways can only be revealed through detailed SFH analysis, rather than radial gradients of mean ages or spectral indices. While mergers and interactions might be the ultimate cause of all radial patterns in quenching, improved simulations are required to confirm this.

    \item We additionally note the importance of accounting for selection effects when identifying rapidly quenching galaxies: in the 11/37 galaxies with synchronous rapid inner and outer quenching, the inner regions were not selected as PSBs due to either no preceding starburst (too weak $\mathrm{H\delta_A}$) or residual star formation / weak AGN (too strong $\mathrm{H\alpha}$). Selection via SFH, rather than spectral indices, would be required to determine the relative fraction of different quenching pathways.

\end{enumerate}

Our results highlight how diverse quenching patterns are in galaxies, and the importance of not oversimplifying into e.g. inside-out or outside-in processes.  
In the future, a full statistical analysis of the different quenching patterns could be obtained through the spatially resolved SFH analysis of all MaNGA galaxies, without the use of simplifying selection techniques such as radial gradients in spectral indices or mean stellar ages. 
Even with high SNR spatially resolved optical spectroscopy, uncovering the causes of the quenching event remains challenging. Details of the molecular gas in these galaxies would help to disentangle the competing effects of past and current gas supplies, and ongoing feedback mechanisms. Hydrodynamic simulations should be useful in interpreting the results, however, detailed physical processes such as cosmic rays and magnetic fields are expected to play an important role in generating the turbulence that is presumably responsible for the decline in star formation efficiency, alongside gravitational disruption by interactions and minor mergers. More development of current generation simulations may be required to reproduce all of the diversity of quenching mechanisms present in the real Universe.

\section*{Acknowledgements}

We thank the Referee for suggestions that improved the paper. We thank Am\'elie Santonge and Andrew Cameron for useful discussions and comments. 
We thank Yingjie Peng for providing data. 
HL and ACC acknowledge support from a UKRI Frontier Research Guarantee Grant (PI Carnall; grant reference EP/Y037065/1). 
VW acknowledges Science and Technologies Facilities Council (STFC) grants ST/V000861/1 and ST/Y00275X/1, and Leverhulme Research Fellowship RF-2024-589/4. 
YMC acknowledges support by the National Natural Science Foundation of China, NSFC Grant Nos. 12333002.

Funding for the Sloan Digital Sky 
Survey IV has been provided by the 
Alfred P. Sloan Foundation, the U.S. 
Department of Energy Office of 
Science, and the Participating 
Institutions. 

SDSS-IV acknowledges support and 
resources from the Center for High 
Performance Computing  at the 
University of Utah. The SDSS 
website is www.sdss4.org.

SDSS-IV is managed by the 
Astrophysical Research Consortium 
for the Participating Institutions 
of the SDSS Collaboration including 
the Brazilian Participation Group, 
the Carnegie Institution for Science, 
Carnegie Mellon University, Center for 
Astrophysics | Harvard \& 
Smithsonian, the Chilean Participation 
Group, the French Participation Group, 
Instituto de Astrof\'isica de 
Canarias, The Johns Hopkins 
University, Kavli Institute for the 
Physics and Mathematics of the 
Universe (IPMU) / University of 
Tokyo, the Korean Participation Group, 
Lawrence Berkeley National Laboratory, 
Leibniz Institut f\"ur Astrophysik 
Potsdam (AIP),  Max-Planck-Institut 
f\"ur Astronomie (MPIA Heidelberg), 
Max-Planck-Institut f\"ur 
Astrophysik (MPA Garching), 
Max-Planck-Institut f\"ur 
Extraterrestrische Physik (MPE), 
National Astronomical Observatories of 
China, New Mexico State University, 
New York University, University of 
Notre Dame, Observat\'ario 
Nacional / MCTI, The Ohio State 
University, Pennsylvania State 
University, Shanghai 
Astronomical Observatory, United 
Kingdom Participation Group, 
Universidad Nacional Aut\'onoma 
de M\'exico, University of Arizona, 
University of Colorado Boulder, 
University of Oxford, University of 
Portsmouth, University of Utah, 
University of Virginia, University 
of Washington, University of 
Wisconsin, Vanderbilt University, 
and Yale University.

\textit{Software:} \textsc{Astropy} \citep{astropy}, \textsc{Bagpipes} \citep{bagpipes2018,bagpipes2019}, \textsc{Celerite2} \citep{celerite,celerite2}, \textsc{Marvin} \citep{marvin}, \textsc{Matplotlib} \citep{matplotlib}, \textsc{MultiNest} \citep{multinest}, \textsc{Nautilus} \citep{nautilus}, \textsc{Numpy} \citep{numpy}, \textsc{pipes\_vis} \citep{pipes_vis}, \textsc{pyMultiNest} \citep{pymultinest}, \textsc{Scipy} \citep{scipy}, \textsc{Seaborn} \citep{seaborn}

For the purpose of open access, the author has applied a Creative Commons Attribution (CC BY) licence to any Author Accepted Manuscript version arising.

\section*{Data Availability}
All utilised MaNGA data are publicly available at the SDSS database \url{https://www.sdss4.org/dr17/} or through \texttt{Marvin} at \url{https://dr17.sdss.org/marvin/}. The stacked spectra and posterior samples of all 50 fitted CPSBs are available at \url{https://doi.org/10.17630/ac0b406c-1c59-41e6-8b73-026790a0c1ca}. Those for all 41 RPSBs are available at \url{https://doi.org/10.17630/e32daf65-6997-4527-9008-b9f1ea6b86e8}. Python scripts to recreate the figures in Section \ref{sec:results} are available at \url{https://github.com/HinLeung622/MaNGA_PSB_evo_pathways_scripts}.



\bibliographystyle{mnras}
\bibliography{biblist} 




\appendix

\section{Fitting central regions of RPSBs with other SFH models} \label{apx:central_sfh}
\begin{figure}
    \centering
    \includegraphics[width=\columnwidth]{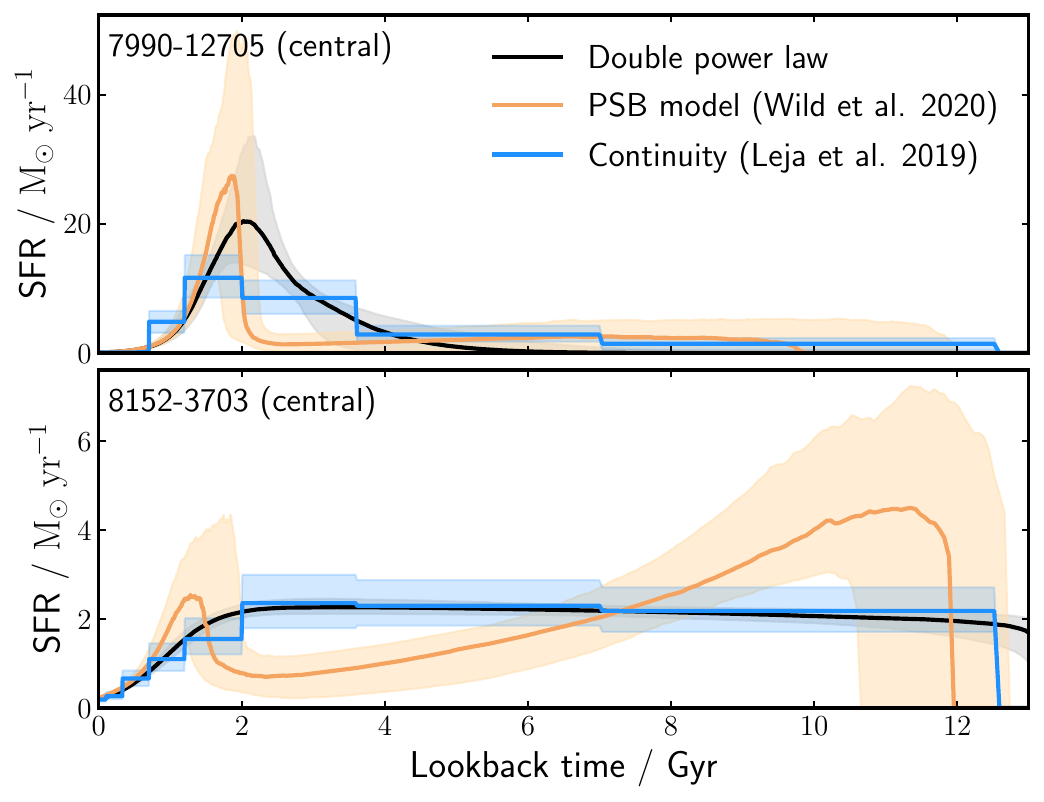}
    \caption{Comparing the fitted SFHs of three SFH models for the central regions of two RPSBs. The Plate-IFU of the galaxy is denoted in the top left of each panel. The SFH models compared are the double power law model (black), the \protect\cite{wild2020} PSB model (orange) and the \protect\cite{leja2019} non-parametric ``continuity'' model (blue). Solid lines mark the median posterior SFH while the shaded regions represent its $1\sigma$ uncertainty. We found the PSB model's fixed $\beta$ prior tend to introduce unwanted rapid rise in the recent starburst of fitted central SFHs, while the other two models reach good agreements in the fitted SFHs of most central regions.}
    \label{fig:compare_sfh_models}
\end{figure}

Aside the fiducial double power law SFH model used for the central regions of RPSBs, we also tested using the same SFH as the PSB regions, the \cite{wild2020} two-component PSB model, and the \cite{leja2019} non-parametric ``continuity'' model. Figure \ref{fig:compare_sfh_models} compares their best fit SFHs for the central regions of two RPSBs. Due to the fixed rapidly rising slope of the recent starburst ($\beta=250$) in the PSB model, the model imposes the central region must have experienced two distinct period of active star formation, when the true SFH could be a more gradual rise during lookback times $2-6\;$Gyr. When we tried relaxing the fixed prior on $\beta$, the higher number of free parameters in the PSB model compared to the double power law model resulted in excessive model flexibility, where the time since the peak of the starburst ($t_\mathrm{burst}$) becomes poorly constrained. Thus, we opt for the simpler double power law model over the PSB model.

Figure \ref{fig:compare_sfh_models} also shows that for most central regions, changing from using the double power law model to the non-parametric ``continuity'' model yields no significant change in the fitted SFHs. This is expected, as both models have been shown to perform well when fitting galaxies with a wide range of true SFHs \citep{carnall2019,leja2019}. However, considering that the double power law model requires less computation per fit and provides a more intuitive comparison to the outer PSB regions' SFHs (e.g. comparing when the recent quenching began), we favour the former model to fit the central regions.

\section{Fitted star-formation histories and metallicity levels}\label{apx:SFH}

Figure \ref{fig:SFH_all_RPSB} shows the derived SFH's and metallicity evolution for the PSB regions of the 41 fitted RPSBs in our sample. The spectral fitting methods used are described in Section~\ref{sec:RPSB_fitting}. SFH parameters such as the time since burst, burst mass fraction, etc are listed in Table \ref{tab:RPSB_PSB_results}. 

\begin{figure}
    \centering
    \includegraphics[width=\columnwidth]{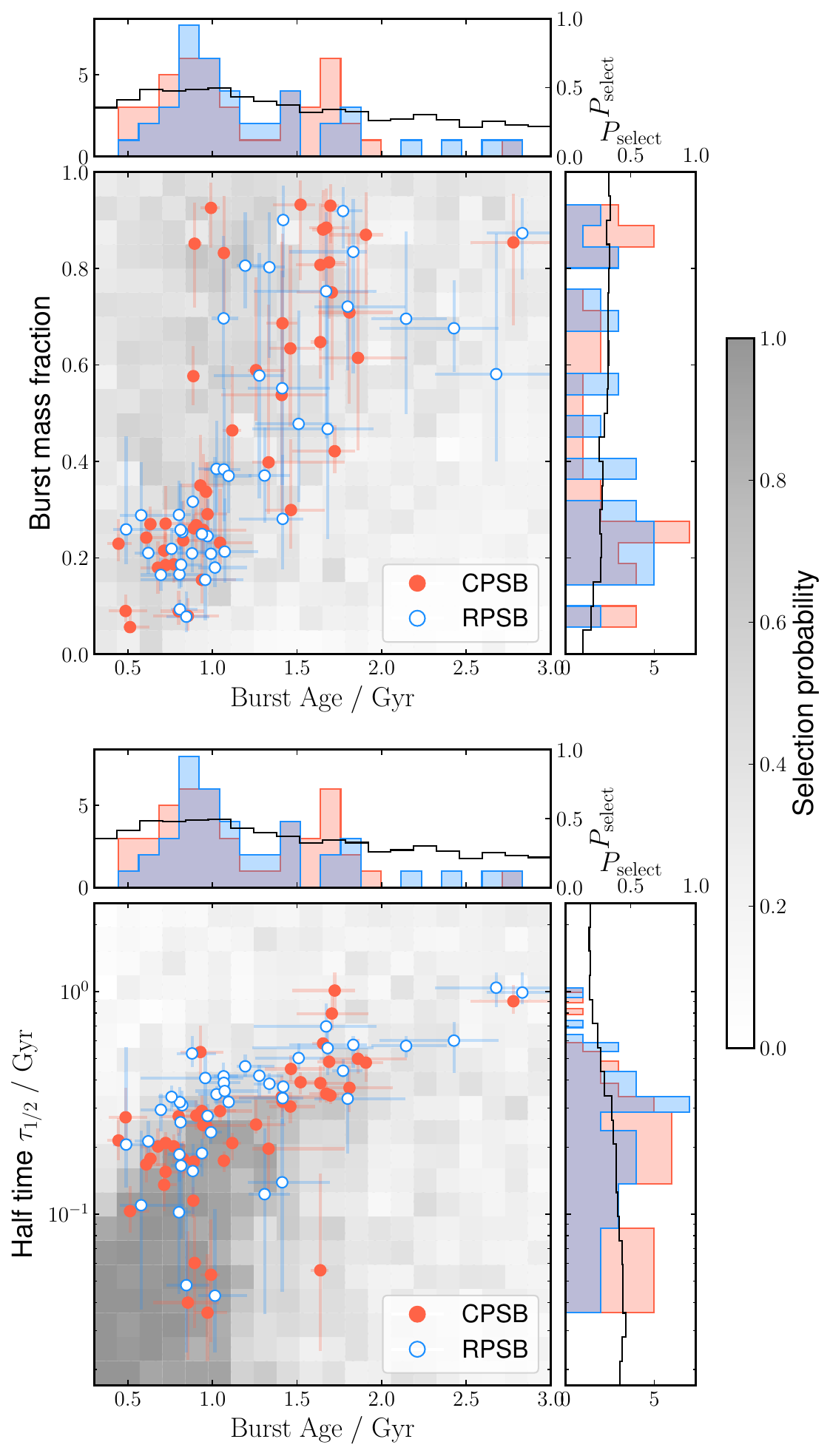}
    \caption{Comparing fitted SFH properties of the PSB regions in CPSBs (red closed dots) and RPSBs (blue open dots). \textbf{Top}: Fitted time since peak of starburst ($t_\mathrm{burst}$) plotted against burst mass fraction ($f_\mathrm{burst}$). \textbf{Bottom}: Fitted time since peak of starburst ($t_\mathrm{burst}$) plotted against the starburst's quenching timescale as measured by its SFH halving timescale ($\tau_{1/2}$). For both panels, histograms that show the 1D distributions of the x- and y-axes parameters for both PSB types are plotted in the side windows. The background greyscale displays the simulated PSB selection probability. The PSB regions in both PSB types are very similarly distributed.}
    \label{fig:CPSBvsRPSB_scatter_SFH_full}
\end{figure}

\begin{figure*}
    \centering
    \includegraphics[width=\textwidth]{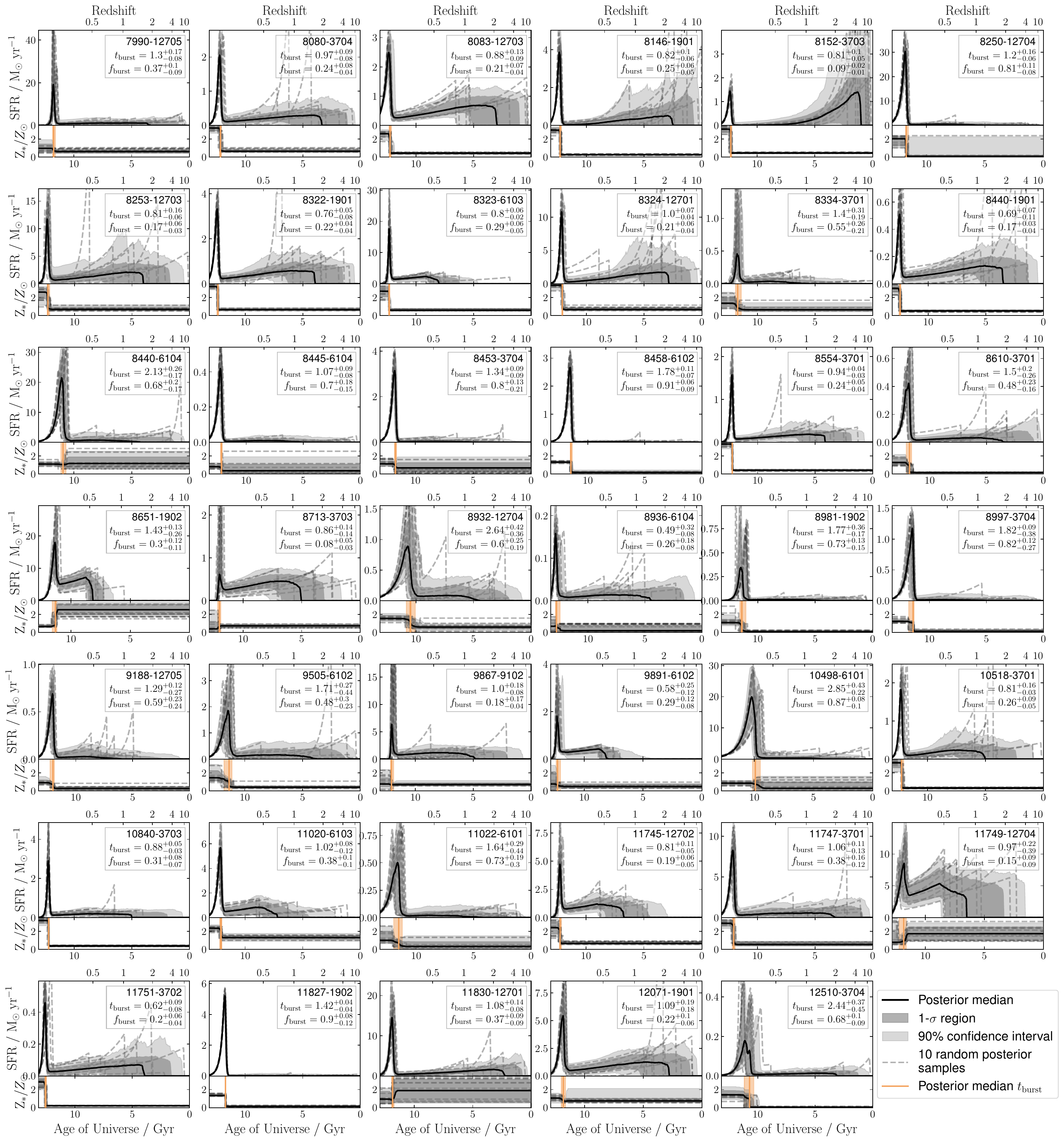}
    \caption{Fitted SFHs (top panel of each subplot) and metallicity evolution (bottom panel of each subplot) of the PSB regions of all 41 successfully fitted RPSBs. The solid black lines, grey and light grey shaded regions mark the posterior median, the $1\sigma$ region and 90\% confidence interval of the fitted SFHs or metallicity histories. Ten random samples drawn from the posterior distributions are shown using grey dashed lines. The vertical orange lines and orange shaded regions mark the posterior median and $1\sigma$ region of the time of peak of starburst ($t_\mathrm{burst}$).}
    \label{fig:SFH_all_RPSB}
\end{figure*}

To consider selection effects introduced by our PSB classification scheme (Section \ref{sec:CPSBvsRPSB_data}), we calculate the theoretical selection fraction given models within our prior space as a function of the parameters of interest. 
First, we create $10^6$ mock galaxy models from the assumed SFH and metallicity priors (the middle section of Table \ref{tab:priors}) using randomly drawn galaxy properties from the prior space. 
We generate two sets of model spectra with \textsc{Bagpipes}: one set with nebular emission lines and one set without. 
The absorption spectral index $\mathrm{H\delta_A}$ is measured from the spectrum without nebular emission, while emission equivalent width $\mathrm{W(H\alpha)}$ is measured by dividing the flux difference between the two spectra within the $\mathrm{H\alpha}$ feature passband by its pseudo-continuum, following \cite{manga_dap}.
The fraction of mock galaxies that were selected through the criteria given in Section \ref{sec:CPSBvsRPSB_data} is taken as the selection probability within our prior.
In Figure \ref{fig:CPSBvsRPSB_scatter_SFH_full}, we show the selection probability as a grey-scale overlaid onto Figure \ref{fig:CPSBvsRPSB_scatter_SFH}.
The averaged 1D functions are projected above and to the right in solid black lines.
We note that the priors are deliberately inclusive of non-PSBs, which results in the low overall selection probability seen.

As indicated by the grey-scale in the top panel of Figure \ref{fig:CPSBvsRPSB_scatter_SFH_full}, PSBs with a more ancient burst ($t_\mathrm{burst}>1\;$Gyr) and weaker burst mass fraction ($f_\mathrm{burst}<0.2$) are unlikely to be detected. 
This is due to a weak $\mathrm{H\delta_A}$ absorption strength for these PSBs, which partially shapes the positive correlation seen between $t_\mathrm{burst}$ and $f_\mathrm{burst}$. 
However, there is relatively little selection effect on PSBs with recent bursts or on PSBs with older bursts with higher burst mass fractions, assuming our priors reflect the local PSB population's SFH properties. 
Comparing the selection probability with the measured trend, there is a significant lack of high-$f_\mathrm{burst}$, low-$t_\mathrm{burst}$ PSBs. We suggest three possible explanations:
\begin{itemize}
    \item High-$f_\mathrm{burst}$ PSBs tend to have high estimation uncertainties in $f_\mathrm{burst}$. It is possible that these PSBs in reality have much more modest $f_\mathrm{burst}>0.5$ values similar to the low-$t_\mathrm{burst}$ PSBs, but their $f_\mathrm{burst}$ estimates were inflated due to uncertainties.
    \item The effects of the burst age-mass degeneracy \citep{wild2007,wild2020} leads to a degeneracy between $t_\mathrm{burst}$ and $f_\mathrm{burst}$ in the same direction with the correlation seen. This degeneracy might have scattered the measured $t_\mathrm{burst}$ and $f_\mathrm{burst}$ values along the degeneracy's direction (high-$t_\mathrm{burst}$-high-$f_\mathrm{burst}$ to low-$t_\mathrm{burst}$-low-$f_\mathrm{burst}$), contributing to the trend observed.
    \item The priors used for estimating the selection probability might not reflect the SFH properties of local PSBs. If most local PSBs tend to experience slower quenching than our prior, a significant fraction of high-$f_\mathrm{burst}$ low-$t_\mathrm{burst}$ PSBs will have low selection probability due to ongoing residual star formation causing high $\mathrm{W(H\alpha)}$. This could explain the lack of high-$f_\mathrm{burst}$, low-$t_\mathrm{burst}$ PSBs in the left panel of Figure \ref{fig:CPSBvsRPSB_scatter_SFH}.
\end{itemize}
The exact origin of the positive correlation requires careful investigation, and we leave to future work.

In the bottom panel of Figure \ref{fig:CPSBvsRPSB_scatter_SFH_full}, the positive correlation between $t_\mathrm{burst}$ and $\tau_{1/2}$ in PSBs can be mostly explained by selection effects as suggested by the background grey scale. 
For a recent burst ($t_\mathrm{burst}<1\;$Gyr), PSBs have a high selection likelihood across a large range of rapid to intermediate quenching timescales, thus the wide area covered by the deeper shade grey scale at this burst age range.
For older bursts, only slower quenching PSBs will have high enough burst mass fraction for them to be selected.
This results in the thinner deeper shade grey scale region at this higher burst age range.
The high selection parameter space can be described as a funnel. 
Most of both PSB samples appears to follow the slower quenching limit of this funnel region. 

Since the correlations between SFH properties shown in Figure \ref{fig:CPSBvsRPSB_scatter_SFH} and \ref{fig:CPSBvsRPSB_scatter_SFH_full} are strongly shaped by selection effects, we do not draw conclusions about the demographics of PSB regions from these results.

\section{Fitted SFHs of RPSBs} \label{apx:RPSB_sfhs}

\begin{figure*}
    \centering
    \includegraphics[width=\textwidth]{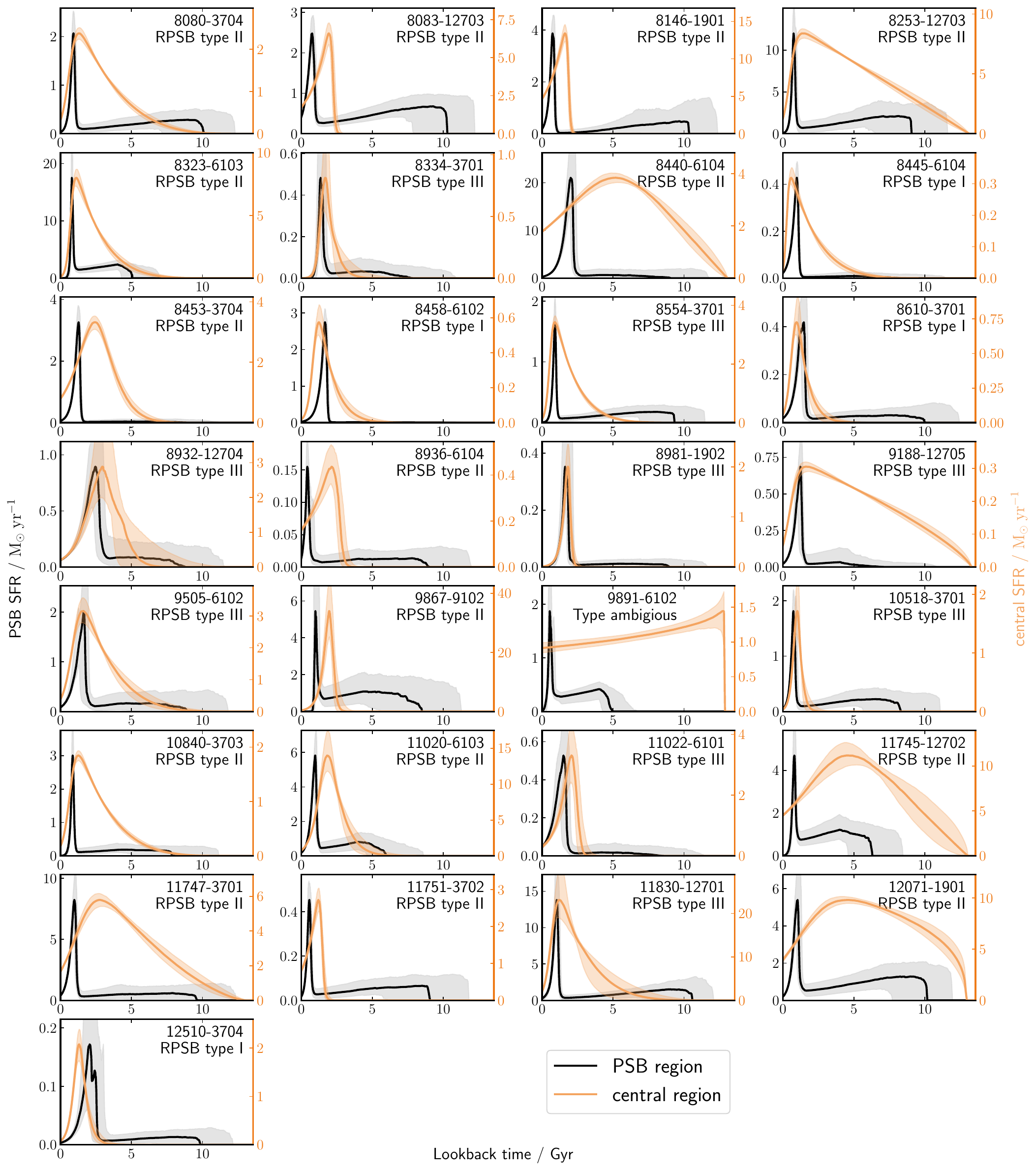}
    \caption{The fitted SFHs of the PSB (black) and central (orange) regions of the remaining 29 RPSBs. All elements of the figure have the same meaning as in Figure \ref{fig:RPSBvsCentral_sfh}.}
    \label{fig:RPSBvsCentral_sfh_apx}
\end{figure*}

\begin{figure*}
    \centering
    \includegraphics[width=\textwidth]{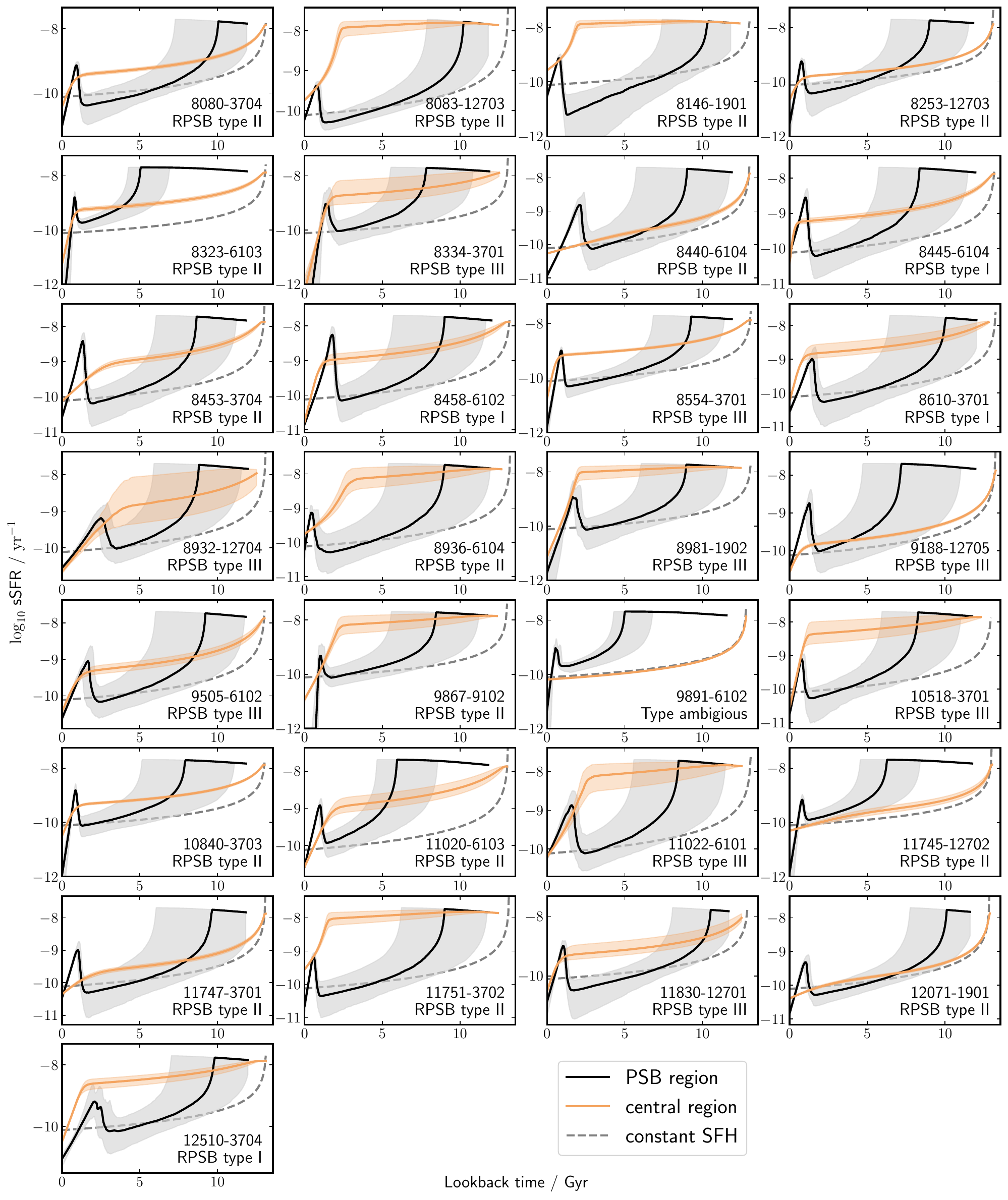}
    \caption{The fitted sSFR of the PSB (black) and central (orange) regions of the remaining 29 RPSBs. All elements of the figure have the same meaning as in Figure \ref{fig:RPSBvsCentral_sfh}.}
    \label{fig:RPSBvsCentral_ssfr_apx}
\end{figure*}

In Section \ref{sec:centralSFH}, we showed the fitted SFHs and sSFR evolution of the outer PSB regions and central regions of 8 example RPSBs in our sample. In this appendix, Figures \ref{fig:RPSBvsCentral_sfh_apx} and \ref{fig:RPSBvsCentral_ssfr_apx} show the fitted SFH and sSFR evolution of all 29 other RPSBs that are not removed due to poor fitted GP noise corrections in their central stacked spectra, in the same style as Figure \ref{fig:RPSBvsCentral_sfh}.

\bsp	
\label{lastpage}
\end{document}